\documentclass[11pt,preprint]{aastex}

\begin{document}
\title{Carbon Enhanced Metal-Poor Stars. \\
I. Chemical Compositions of 26 Stars\footnote{Based on data collected
at the Subaru Telescope, which is operated by the National
Astronomical Observatory of Japan.}}

\author{Wako Aoki\altaffilmark{2}, 
Timothy C. Beers\altaffilmark{3}, Norbert Christlieb\altaffilmark{4},
John E. Norris\altaffilmark{5}, Sean G. Ryan\altaffilmark{6,7}, Stelios
Tsangarides\altaffilmark{6}}

\altaffiltext{2}{National Astronomical Observatory, Mitaka, Tokyo,
181-8588, Japan; email: aoki.wako@nao.ac.jp}
\altaffiltext{3}{Dept. of Physics \& Astronomy, CSCE: Center for the Study of
Cosmic Evolution, and JINA: Joint Institute for Nuclear Astrophysics,
Michigan State University, E. Lansing, MI
48824; beers@pa.msu.edu}
\altaffiltext{4}{Hamburger Sternwarte, University of Hamburg, Gojenbergsweg
  112, D-21029 Hamburg, Germany; nchristlieb@hs.uni-hamburg.de}
\altaffiltext{5}{Research School of Astronomy and Astrophysics, The
Australian National University, Mount Stromlo Observatory, Cotter
Road, Weston, ACT 2611, Australia; email: jen@mso.anu.edu.au}
\altaffiltext{6}{Department of Physics and Astronomy, The Open
University, Walton Hall, Milton Keynes, MK7 6AA, UK; email:
stsangarides@gmail.com}
\altaffiltext{7}{Present address: Centre for Astrophysics Research, STRI and School of Physics,
Astronomy and Mathematics, University of Hertfordshire, College Lane,
Hatfield AL10 9AB, United Kingdom; s.g.ryan@herts.ac.uk} 

\begin{abstract} 

The chemical compositions of 26 metal-poor stars that exhibit strong
CH and/or C$_{2}$ molecular bands are determined based on
high-resolution spectroscopy. We define carbon-enhanced stars taking
account of the carbon abundance ratio ([C/Fe]) and the evolutionary
status, which is a slight modification over previous
definitions. Twenty two stars in our sample satisfy our modified
definition for Carbon-Enhanced Metal-Poor (CEMP) stars. In addition,
we measure Na abundances for nine other carbon-enhanced stars for
which abundances of other elements have been previously
reported. Combining our new sample with the results of previous work,
we investigate the abundance and evolutionary status of a total of 64
CEMP stars. The following results are obtained: (1) All but one of
the 37 stars with [Fe/H] $\geq -2.6$ exhibit large excesses of barium
([Ba/Fe] $> +0.5$), while the other 27 stars with lower metallicity
exhibit a large scatter in their barium abundance ratios ($-1.2 <$
[Ba/Fe]$ < $+3.3). (2) A correlation between the carbon and barium
abundance ratios ([C/Fe] and [Ba/Fe]) is found in Ba-enhanced objects
(comprising 54 stars), suggesting that the origin of the observed
carbon excess in Ba-enhanced stars is nucleosynthesis in asymptotic
giant branch (AGB) stars, where the main $s$-process occurs. The
correlation between the barium abundance ratio and that of carbon plus
nitrogen ([(C+N)/Fe]) is relatively weak, because of the large
excesses of nitrogen in some extremely metal-poor stars. (3) The
majority of the Ba-enhanced stars have $-1.0 < $ [C/H] $ < 0.0$, and a
clear cutoff exists at [C/H] $\sim 0$, which we take as the limit of
carbon-enrichment by metal-poor AGB stars. Within the above range, the
[C/H] of the Ba-enhanced CEMP stars decreases, on average, by up to
0.6~dex from the main-sequence turnoff up the red-giant branch,
suggesting some dilution of carbon enhancement during their
evolution. The [C/H] values of Ba-normal stars are relatively low,
with a wide distribution. (4) The difference in the distributions of
evolutionary status between Ba-enhanced and Ba-normal CEMP stars
suggested by our previous work is not statistically confirmed by the
present, enlarged sample. (5) Excesses of Na are found in stars with
extremely large enhancements of C, N and Ba, suggesting efficient
production of this element by AGB nucleosynthesis. The implications
of these results on the origins of carbon in CEMP stars, in particular
for Ba-normal stars, are discussed.


\end{abstract} 

\keywords{ 
nuclear reactions, nucleosynthesis, abundances --- stars: abundances
--- stars: AGB and post-AGB ---  stars: carbon --- stars: Population II}

\section{Introduction}\label{sec:intro}

Large modern surveys of metal-deficient stars in the Galaxy have revealed that
the fraction of carbon-enhanced objects is significant at low metallicity
\citep{beers92, christlieb03}.  The fraction appears to increase with
decreasing metallicity (Beers \& Christlieb 2005; Lucatello et al. 2006),
although the derived fraction depends somewhat on one's choice of the definition
of the carbon-enhancement phenomenon. Many years ago, \citet{keenan42} coined
the term ``CH stars'' to refer to ``high-velocity'' carbon stars that exhibit
very strong G-bands due to the CH molecule, and otherwise weak lines of other
metals usually used to classify stellar spectra. The small sample of stars he
had available at that time had spectra that suggested high luminosities, i.e.,
membership on the giant branch. The nomenclature was extended to objects of
lower luminosity by
\cite{bond74}, who found subgiants with similar elemental compositions. More
recently, surveys for carbon stars at high Galactic latitude have also been made
\citep{green94, totten98,totten00,christlieb01}. The Carbon-Enhanced Metal-Poor
(CEMP) stars found by recent surveys of metal-poor stars may well be closely
related to these objects.

An important question concerning the nature of CEMP stars is that of
the astrophysical origin of the carbon excess that is observed in
these objects. One established scenario is carbon production by
nucleosynthesis in an asymptotic giant branch (AGB) star. Since most
CEMP stars that are observed in the present Galaxy are too old to be
intrinsic AGB stars\footnote{There exists one recently-discovered
exception, CS~30322--023, which \citet{masseron06} argue is likely to
be an extremely metal-poor AGB star that is presently undergoing
thermal pulses. This star also appears in our present sample.}, it
is usually assumed that the AGB star responsible for carbon production
was once the primary of a binary system, and the elements produced
during its lifetime and dredged to its outer atmosphere were
transferred to its companion prior to the primary evolving to become a
faint white dwarf. Observational support for this scenario is found
from the observed excess of $s$-process elements, as well as carbon,
in classical CH stars \citep[e.g. ][]{wallerstein64, vanture92} and
some CEMP stars \citep[e.g., ][]{norris97a}. Other evidence comes from
the binarity of the objects found in some CH and CEMP stars identified
from radial-velocity monitoring \citep{mcclure84,
mcclure90,preston01}. The fraction of binary stars among CEMP stars
has been recently studied by \citet{lucatello05}, who concluded that
all CEMP stars with s-process element overabundances (which Beers \&
Christlieb 2005 refer to as CEMP-s stars) could statistically belong
to binary systems, given the limited observational constraints on the
radial-velocity variations observed to date.\footnote{Contributions of
the s- and r- processes are estimated from the abundance ratios of
neutron-capture elements such as [Ba/Fe], [Eu/Fe], and [Ba/Eu]. For
instance, the term 's-process element-enhanced stars' indicates
objects that have high [Ba/Fe]($ > +0.5$) and high [Ba/Eu] ($\sim +1$).}

However, in some CEMP stars, no excess of neutron-capture elements is
found \citep[e.g., ][]{norris97b,aoki02b}; these stars are referred to
as CEMP-no stars \citep{beers05}. At least one CEMP star
\citep[CS~22892--052; ][]{sneden96} has been reported with an excess
of neutron-capture elements associated with the r-process. Lucatello
et al. (in preparation) has identified a few other such stars, hence
the class CEMP-r is no longer based on a single object. There are a
growing number of CEMP stars with apparent contributions from {\it
both} the r-process and the s-process, the CEMP-r/s stars (see Beers
\& Christlieb 2005 for further discussion of these classes). The
existence of such a wide variety of elemental abundance patterns
associated with CEMP stars strongly suggests that other mechanisms may
be responsible for their nucleosynthetic history beyond those
involving AGB stars.  \citet{ryan05} compiled the elemental and
isotopic abundances of 19 CEMP stars measured by previous work where
the $^{13}$C/$^{12}$C ratio was studied, and considered possible
differences between Ba-enhanced CEMP stars (CEMP-s) and Ba-normal ones
(CEMP-no). This discussion was necessarily somewhat limited because of
the small sample size.

Our long-running investigation of these interesting objects has led us
to undertake a high-resolution spectroscopic study for an expanded
sample of candidate CEMP stars. We have thus far obtained spectra for
26 stars with the Subaru Telescope High Dispersion Spectrograph
\citep{noguchi02}; we report on their chemical compositions in this
paper. We also analyze redder spectra, in order to measure their Na
abundances, for another nine stars for which abundances of other elements
have been reported by previous work. Based on the results of our
present abundance study and on previous work, we discuss the
distribution of the carbon-enhancement, the fraction of Ba-enhanced
stars, and the metallicity distribution among CEMP stars. The
evolutionary status of our sample is taken into consideration in our
discussion. The sample selection, observations, and data-reduction
procedures are reported in \S 2. Details of the abundance analyses are
presented in \S 3. Our modified definition of carbon-enhanced objects is given
in \S 4. We discuss the abundance distribution of barium and the
differences between the Ba-enhanced and Ba-normal stars in \S 5. In \S
6 we present our conclusions.

\section{Observations and Measurements}

\subsection{Sample Selection and Photometry}

Stars in our main program were selected on the basis of
medium-resolution spectroscopic follow-up of metal-poor candidates
from the HK survey \citep{beers92,beers99} and the Hamburg/ESO survey
(HES; Christlieb 2003) that exhibited apparently strong CH G bands and
metallicities of [Fe/H] $< -2.0$.\footnote{[A/B] = $\log(N_{\rm
A}/N_{\rm B})- \log(N_{\rm A}/N_{\rm B})_{\odot}$, and
$\log\epsilon_{\rm A} = \log(N_{\rm A}/N_{\rm H})+12$ for elements A
and B.} The majority of the HES stars considered in the present paper
were drawn from the list of carbon-enhanced stars of Christlieb et
al. (2001), which sought to identify stars on the basis of their
carbon enhancement, rather than low metallicity. The list of objects
is provided in Table~\ref{tab:obs}, where coordinates and details of
the observations are given. We note that HE~0039--2635 is identical to
CS~29497--034, which was studied in detail by \citet{barbuy97},
\citet{hill00}, and \citet{barbuy05}. This star was re-identified as a
candidate CEMP star by the HES.

Effective temperatures for the stars in our sample are estimated from
photometric data using the color-temperature relation provided by
\citet{alonso96} and \citet{alonso99} (see \S
\ref{sec:param}). Optical $BVRI$ photometry (Johnson-Kron-Cousins) for
these stars are reported by Beers et al. (2006). The results
are listed in Table~\ref{tab:photo}. Errors in the photometric
measurements are typically on the order of 0.01--0.02~mag. Near
infrared $JHK$ photometry was collected from the database of the Two
Micron All Sky Survey (2MASS) Point Source Catalog
\citep{skrutskie06}. Table~\ref{tab:photo} also lists our adopted
estimates of interstellar reddening ($E(B-V)$), which are total
line-of-sight values derived from the dust map of
\citet{schlegel98}. The extinction in each band is obtained from the
reddening relation provided in their Table 6.

\subsection{High-Resolution Spectroscopy}

High-resolution spectroscopic observations for the stars in our sample was
obtained with Subaru/HDS, with a spectral resolving power $R = 50,000$ (a slit
width of 0.72~arcsec), in October 2002 and May 2003. The atmospheric dispersion
corrector (ADC) was used in all observing runs. Two EEV-CCDs were used with two by
two on-chip binning, providing approximately a three-pixel sampling of the resolution
element. The spectra obtained in October 2002 cover the wavelength range
4100--6850 {\AA}, with a small gap in the coverage between 5440 and 5520~{\AA}
due to the physical separation between the two CCDs. The wavelength coverage was
slightly ($\sim 70$~{\AA}) shifted to shorter wavelengths in the May 2003 run.
For CS~29528--028, which is the hottest star in our sample, a blue spectrum
(3550--5250~{\AA}) was also obtained during the October 2002 run. 

Spectra of several stars not listed in Table~\ref{tab:obs} were also obtained.
These stars, however, turned out to exhibit severe molecular absorption (C$_{2}$
and CN), and special care is required for their abundance analysis. These
spectra will be reported on separately \citep{goswami06}.

\subsection{Red Spectra of Nine Additional Stars}

Table~\ref{tab:naobs} lists the object names and observing log for stars for
which we have also provided an analysis of Na. All stars other than
CS~22183--015 were studied by our previous work \citep{aoki02c, aoki02d}, based
on blue spectra. CS~22183--015 was studied by \citet{johnson02} and
\citet{cohen06}, but the Na abundance was not measured by these authors. 

Most of the red spectra were obtained with Subaru/HDS in 2005 during scheduled
service observing. The single exception is CS~22898--027, which was observed
during the HDS commissioning phase in 2000. Photon counts per 0.18~km~s$^{-1}$
pixel at 5900~{\AA} are listed in Table~\ref{tab:naobs}. Note that although some objects
were observed for the purpose of radial-velocity monitoring, the counts are
sufficient for our present analysis.

\subsection{Data Reduction}

Standard data reduction procedures (bias subtraction, flat-fielding, background
subtraction, extraction, and wavelength calibration) were carried out with the
IRAF echelle package\footnote{IRAF is distributed by the National Optical
Astronomy Observatories, which is operated by the Association of Universities
for Research in Astronomy, Inc. under cooperative agreement with the National
Science Foundation.}. Suspected cosmic-ray hits are removed using the technique
described by \citet{aoki05}.

The sixth column of Table~\ref{tab:obs} lists the photon count (per
1.8~km~s$^{-1}$ pixel) collected at 5200~{\AA}. Since the primary purpose of
this work is to investigate the abundance characteristics for a large sample, we
pursued only moderate signal-to-noise (S/N) ratios that could be obtained with
relatively short integrations. Note also that the data quality is not uniform in
our sample. One reason for this is that complete optical photometry was not
available at the time of the spectroscopic observing run. Moreover, our sample
consists of stars having quite different colors. These factors made the
estimation of exposure times required to obtain uniform S/N ratios quite difficult.

\subsection{Measurements of Equivalent Widths}

Equivalent widths are measured for isolated atomic lines by fitting gaussian
profiles, while a spectrum-synthesis technique is applied to molecular bands.
The measured equivalent widths of a number of elements are listed in
Tables~\ref{tab:ew1}, \ref{tab:ew2}, and \ref{tab:ew3}. We selected 10 elements
(12 species) for the measurements of equivalent widths. 

Table~\ref{tab:ew4} lists the equivalent widths of lines in the blue range for
CS~29528--028. As shown below, the heavy neutron-capture elements in this star
are significantly overabundant, and equivalent widths of the Sr, Zr, La, Ce, and
Nd lines are measured.  

The equivalent widths of Na lines for the nine other stars are listed in
Table~\ref{tab:ew5}. The Na~D lines could not be measured for CS~31062--050
because of severe contamination from interstellar Na absorption. The Na
abundance of this object is determined from the $\lambda$5688\,{\AA} line.

\subsection{Measurements of Radial Velocities}

We measure heliocentric radial velocities ($V_{\rm r}$) using isolated
\ion{Fe}{1} lines. The results are listed in Table~\ref{tab:obs}. The
standard error ($\sigma / \sqrt{n}$, where $\sigma$ is the standard
deviation of the values from individual lines and $n$ is the number of
lines used) is adopted as the random error of the measurements reported in
the table.  The wavelength calibration was made using Th-Ar comparison
spectra that were obtained during the observing runs.  Possible
systematic errors of the wavelength calibration due to the stability
of the spectrograph are no more than 0.4~km~s$^{-1}$ \citep[see
][]{aoki05}. Combining this possible systematic error with the random
errors (3 $\sigma$), the uncertainties of the reported radial velocity
measurements are 0.5--1.0~km~s$^{-1}$.

The $V_{\rm r}$ of BS~16929--005 measured in the present work agrees
with that of \citet{honda04a} ($-51.29$~km~s$^{-1}$) within the stated
errors. No significant variation was found for this object between the
two observations, taken roughly two years apart. The radial velocities and the
binarity of CS~22948--027 and CS~29497--034 were investigated by
\citet{barbuy05}. Our results agree, within the measurement errors,
with the predictions from the orbital parameters for these two stars
derived by these authors. Different orbital parameters for CS~22948--027 were
determined by \citet{preston01}. Although the values of the orbital
parameters are quite different between the two studies, both predict a similar
velocity at the epoch of our observation. Hence, our
measurement for a single epoch does not provide meaningful constraints
on the determination of the orbital parameters for CS~22948--027 at
present. Further monitoring of the radial velocity of this star is strongly
desired.

\section{Chemical Abundance Analyses and Results}\label{sec:ana}

A standard LTE analysis, employing the model atmospheres of \citet{kurucz93}, is
performed for the measured equivalent widths for most of the detectable
elements, while a spectrum-synthesis technique is applied for the CH, C$_{2}$,
and CN molecular bands.

Since the calculations of model atmospheres used in the present analyses are
made for scaled solar abundances, the effect of the enhancements of carbon and
nitrogen on thermal structures is not included. This may have some affect on
cool stars with large excesses of these elements. \citet{hill00} showed that the
temperature at the layers where weak absorption lines form ($\log \tau = -1$ to
0) is a few hundred kelvin higher in the CN-enhanced ([C/Fe] = [N/Fe] = +2 for
[Fe/H]$=-3$) model for $T_{\rm eff}$=4500~K than in the standard model (that is,
the model based on scaled solar abundances). The effect of the enhancements of
carbon and nitrogen on, for instance, iron abundances derived from Fe {\small I}
lines, might be 0.2--0.3~dex, which is estimated from the sensitivity of the
derived abundances to the effective temperatures adopted for a few cool stars
($T_{\rm eff}\lesssim 4500$~K) in our sample (e.g., HE~1447+0102). We note that
the carbon enhancement of CS~30322--023, the coolest object in our sample, is
relatively small ([C/Fe]$=+0.56$, see \S~\ref{sec:cn}), thus the effect of the
carbon enhancement on the thermal structure of the model atmosphere for this
star might be less significant.

\subsection{Atmospheric Parameters}\label{sec:param}

Effective temperatures are often determined in high-resolution spectroscopic
studies by ensuring that the derived abundance of some species (e.g., Fe) is
independent of the excitation potential of the spectral lines employed.
However, since the measurement accuracy of the weak absorption lines is not so
high in our moderate S/N spectra, the spectroscopic method to determine
effective temperature is less than ideal. Instead, we estimate effective
temperatures of our program stars from broad-band photometry, although we note
that the determination of effective temperatures for CEMP stars from photometric
data can also present a challenge, due to the presence of severe molecular
absorption in some photometric bands. 

We first estimate the effective temperatures for the stars in our
sample from their $V-K$, $V-R$, $V-I$, and $R-I$ colors, using the
empirical temperature scale of \citet{alonso96} for CS~29503--010 and
CS~29528--028 (warm stars) and \citet{alonso99} for the others, after
reddening corrections were applied
(Table~\ref{tab:photo}). Corrections for the filter systems which we
employed, in order to put them onto the Alonso et al. scale, were as
reported by \citet{aoki05}. The metallicity applied to the effective
temperature scale is first assumed to be [Fe/H] $= -2.5$, and is then
corrected after the first-pass abundance analysis has been
completed. The determination of the effective temperature and
abundance analysis are iterated until the derived iron abundance is no
longer significantly different from the assumed value. We give
priority to the results from $V-K$ ($T_{\rm eff}(V-K)$), as mentioned
below, which are listed in Table~\ref{tab:photo}. The average of the
effective temperatures from the other three colors is calculated, and
the differences of this average with respect to $T_{\rm eff}(V-K)$ are
also listed in the Table. A typical error of the $V-K$ color is 0.03
mag (errors of the $K$-band photometry and reddening estimates are
usually dominant). The error of the $(V-K)_{0}$ of 0.03 mag results in
an uncertainty in $T_{\rm eff}$ of 90~K.

The effect of molecular absorption is relatively small in the $V$ and $K$ bands
for most stars in our sample. Indeed, the $V-K$ colors calculated by models
including enhancements of carbon and nitrogen by \citet{hill00} are in
reasonable agreement with the values derived by the empirical color--$T_{\rm
eff}$ relation by \citet{alonso99}. For this reason the results from the $V-K$
color is adopted for most of our program stars. However, we made corrections to
the effective temperatures of the following five stars:

(1) {\bf CS~22948--027 and CS~29497--034}:  These stars have been studied in previous work
by \citet{hill00} and \citet{barbuy05}, who adopted $T_{\rm
eff}=4800$~K for both objects. The effective temperatures
we derived from the $V-K$ colors are 330~K and 230~K higher than their
values, respectively. These differences could be due to the differences in the
reddening corrections adopted. Moreover, other colors yield lower
temperatures. Including these results, we adopt 5000~K and 4900~K for
CS~22948--027 and CS~29497--034, respectively, in our analysis.

(2) {\bf HE~1005--1439 and HE~1447+0102}: A large discrepancy between $T_{\rm
eff}(V-K)$ and those estimated from the other three color indices is found in
stars with strong C$_{2}$ and CN molecular bands; in general, the temperature
derived from $V-K$ is lower (the effect of molecular bands on $R-I$ color is
significant), as was also found for CS~22948--027 and CS~29497--034. However, the
opposite results are obtained for HE~1005--1439 and HE~1447+0102, for which
quite high temperatures are derived from the $V-K$ colors. We suspect there is
some problem in the estimate of the temperatures from this color for these
objects, and adopt the lower values.

(3) {\bf CS~29528--028}: The effective temperature of CS~29528--028
from the $V-K$ and other colors is higher than the values found in
most low-metallicity main-sequence turn-off stars. In such stars,
the effect of molecular absorption is almost negligible in the
broad-band photometry, and the $B-V$ color also provides useful
information. The effective temperature of CS~29528--028 from the $B-V$
calibration \citep{alonso96} is 6580~K, significantly lower than that
from $V-K$. We adopted 6800~K for this object.

For the above five stars we assumed larger uncertainties in the estimate of
errors of abundance determination due to the uncertainties in their adopted
atmospheric parameters (see \S~\ref{sec:err}).

We perform an abundance analyses of our program stars in the standard manner for
the measured equivalent widths of Fe {\small I} and Fe {\small II} lines. The
microturbulent velocity ($v_{\rm tur}$) is determined from the Fe {\small I}
lines by demanding that there exist no dependence of the derived abundance on
equivalent widths. For HE~0507--1653, HE~1157--0518, HE~1319--1935, and
HE~2221--0453, however, no or only a few weak Fe {\small I} lines were
available, hence reliable microturbulent velocities could not be directly
determined. In these cases we assumed a microturbulent velocity of
2.0~km~s$^{-1}$, close to the average of the values of the other CEMP giants in
our sample. A larger uncertainty on the microturbulent velocity is assumed for
these three stars in calculations of their abundance errors (\S~\ref{sec:err}).

\begin{figure} 
\includegraphics[width=8.5cm]{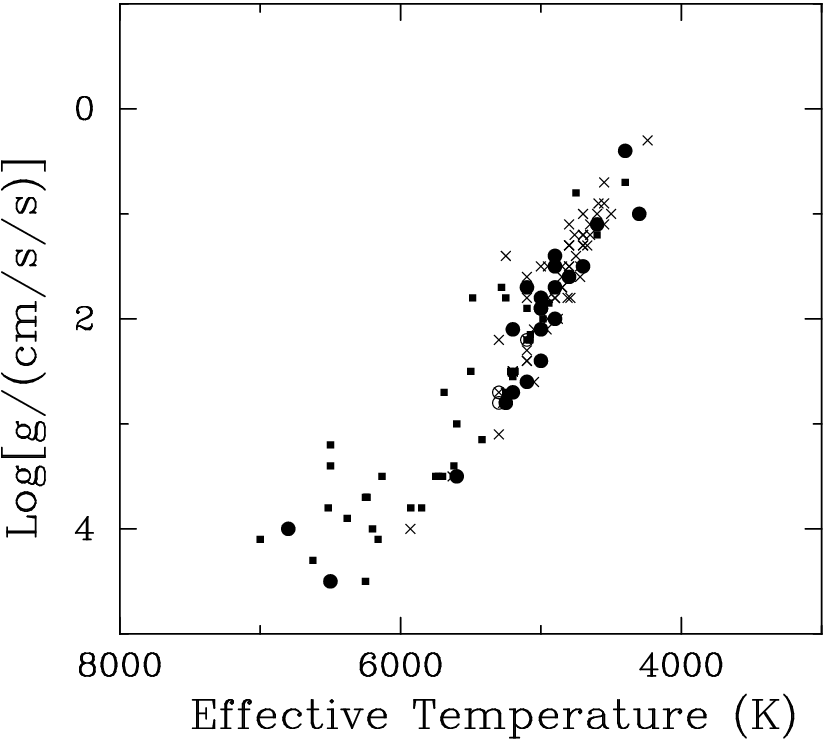} 
\caption[]{The surface gravity log g (in cgs units) determined by analyses of
  iron lines as a function of effective temperature. The filled
  circles indicate stars from our sample, while filled squares and crosses
  indicate other CEMP stars and non carbon-enhanced objects, respectively,
  studied by previous works (see text). Note the absence of
  horizontal-branch stars; if present, red horizontal-branch stars
  would be seen at log g $\sim 2.0$.}

\label{fig:teffg}
\end{figure}

Surface gravities ($\log g$) are determined from the ionization balance between
Fe {\small I} and Fe {\small II}. The final atmospheric parameters are reported
in Table~\ref{tab:param}. Figure \ref{fig:teffg} shows the correlation between
the effective temperatures and gravities for our sample (filled circles) and
also for other stars studied by previous work (see \S \ref{sec:sample} for
references). A clear correlation is found for stars with $T_{\rm eff}<5500$~K,
as most stars in our sample lie along the halo red-giant branch. Our sample also
includes a number of main-sequence turn-off and subgiant stars.

We investigated the correlation between the lower excitation potential
and the derived Fe abundances from individual \ion{Fe}{1} lines. We
found that eight objects among the 26 stars show statistically
significant dependence of the derived abundances on the excitation
potential. The dependence disappears by assuming lower effective
temperatures by 150--250~K than derived from the colors. This result
possibly implies that our calibration based on color indices
systematically overestimates the effective temperature of CEMP
stars. However, given the agreement of the results for similar number
of objects for which sufficient number of \ion{Fe}{1} lines are
measured, the systematic error is unlikely to be as large as 200~K. It
should be noted that the five objects to which we applied special
corrections of effective temperatures show no significant correlation
between the excitation potential and the abundances from individual
\ion{Fe}{1} lines.

For the stars for which Na abundance measurements are made, we adopt the
atmospheric parameters determined by \citet{aoki02c}, \citet{cohen06}, and
\citet{aoki02d} for CS~22957--027, for CS~22183--015, and for the other seven
stars, respectively.

\subsection{Carbon and Nitrogen Abundances}\label{sec:cn}

For the CH 4323~{\AA} band and/or the C$_{2}$ Swan bands, carbon
abundances are determined using a spectrum-synthesis approach. The CH
band is used for stars that have relatively weak absorption features
of molecules due to their high temperature and/or small excesses of
carbon. The C$_{2}$ Swan 0--0 band (5167~{\AA}) is applied to the
analysis for stars in which the CH band is highly saturated. For cool
objects with significant enhancements of carbon, even the C$_{2}$ Swan
0--0 band is severely saturated. In these instances the C$_{2}$ Swan
0--1 band at 5635~{\AA} is used. The results are listed in
Table~\ref{tab:res}, where the absorption band used for the analysis
is also presented for each star. There exist some uncertainties in the
transition probabilities of these molecular bands, which may possibly
cause systematic errors in the carbon abundance determinations. The
abundance analysis for the CEMP subgiant LP~625--44 by
\citet{aoki02a}, using these same three bands, and for which the same
line lists as those used in the present study were adopted,
demonstrated that the derived carbon abundance from the C$_{2}$ bands
is 0.2~dex higher than that from the CH band. Thus, a small systematic
error in the carbon abundance determination possibly exists, depending
on the molecular bands applied to the analysis.

The carbon isotope ratio ($^{12}$C/$^{13}$C) is measured using the C$_{2}$ Swan
1--0 bands at 4736~{\AA} ($^{12}$C$_{2}$) and 4745~{\AA} ($^{12}$C$^{13}$C) for
12 objects in which these bands are detected. The line list and analysis
technique used by \citet{aoki97} are adopted. The results are listed in
Table~\ref{tab:ciso}. Most stars have $^{12}$C/$^{13}$C ratios around 10, a
typical value found in classical CH stars \citep[e.g.,][]{aoki97}, while quite
high values of this ratio are found for three of our program stars.

The nitrogen abundances are estimated from the CN~4215~{\AA} band. For some of
our stars the band appears clearly, and without severe saturation. In these
cases, a reliable fit of synthetic spectra to the observed one is obtained.
However, this band is not detected in seven of our program stars, and only upper
limits on nitrogen abundances are determined (see Table~\ref{tab:res}). For some
cool objects the CN band is so severely saturated that the fitting of the
synthetic spectra is quite uncertain. We compare the observed spectrum and the
synthetic ones calculated changing the nitrogen abundances, and estimate the
fitting errors by eye. In the worst case the uncertainty reaches 0.5~dex.

\subsection{Other Elements}\label{sec:naba}

The abundances of other elements are measured by standard analyses for the
measured equivalent widths. The effects of hyperfine splitting on abundance
determination are significant in the analysis of Ba lines if the abundances of
isotopes with odd mass numbers ($^{135}$Ba and $^{137}$Ba) are high
\citep{mcwilliam98}. This is the case when the contribution of the
r-process is large, while the s-process mostly yields
$^{138}$Ba. Hence, the derived abundances are dependent on the
processes of Ba production that are assumed in the analysis.

We first determine the Ba abundances with a single-line approximation, and
classify our sample into ``Ba-enhanced'' and ``Ba-normal'' stars. Here, a
Ba-enhanced star is defined as having [Ba/Fe] $> +0.5$ (see \S~\ref{sec:ba}). For
Ba-enhanced CEMP stars, we assume that the origin of barium is the s-process,
and re-calculate the abundances including hyperfine splitting, assuming the Ba
isotope ratios estimated by \citet{arlandini99}.\footnote{This assumption is
invalid for CEMP stars with excesses of r-process elements like CS~22892--052
\citep{sneden96}. Such objects are, however, known to be very rare \citep[e.g.,
][]{beers05}, although the abundances of r-process elements are not determined
in the present analysis.} The effect of hyperfine splitting on the abundances
derived from the two resonance lines is on the order of 0.1~dex or smaller. For
Ba-normal stars, we assume the Ba isotope ratios of the r-process component in
solar-system material, and derive the final abundance including the effects of
hyperfine splitting. The result of including hyperfine splitting is
0.1--0.2~dex, depending on the strengths of the Ba lines.

For CS~29528--028, for which a blue spectrum was obtained in the present study,
the abundances of neutron-capture elements other than Ba are measured
(Table~\ref{tab:heavy}). The effect of hyperfine splitting is included in the
analysis of La lines, using the line data by \citet{lawler01}. The
overabundances of the heavy neutron-capture elements (Ba--Nd) are extremely
high, while the excesses of the light neutron-capture elements (Sr and Zr) are
relatively small. This abundance pattern is similar to those found in other CEMP
stars with excesses of s-process elements (e.g., Aoki et al. 2002d).

Sodium abundances are determined for 30 stars, including nine objects for which
abundances of other elements were previously studied, using the D lines
($\lambda\lambda$5889~{\AA}, 5895~{\AA}) for most of the stars with available
red spectra. These lines are known to be significantly affected by non-LTE
effects. The abundances derived from these lines are corrected for these effects
using the results of \citet{takeda03}, who provide the non-LTE corrections for
the 5895~{\AA} line in their Figure 2. We assume the same corrections apply to the
5889~{\AA} line.

Table~\ref{tab:nares} presents the results of the Na abundance measurements. The
results of an LTE analysis are listed in the third column, while the non-LTE
corrections applied to the results from the D lines, which are dependent on the
line strength, are listed in the fourth column. For several stars, the
subordinate lines at 5682 {\AA} and 5688 {\AA} are used to measure the Na
abundances. The non-LTE effects on these weak lines are estimated to be quite
small \citep{takeda03}. The final results are obtained from the non-LTE
corrected Na abundances from the D lines and LTE results from the subordinate
lines, and are listed in the fifth and sixth columns of Table~\ref{tab:nares}.
The Na abundances of four stars are determined using all four of these lines.
For CS~29528--028 and HE~0400-2030, the non-LTE corrected values derived from
the D lines agree well with the results from the subordinate lines. However, for
CS~22957--027, the results from the D lines are higher by 0.5~dex than those
from the other lines, even though a large non-LTE correction (0.5~dex) is
applied to the results from the D lines. By contrast, for HE~0507--1653, the
results from the subordinate lines are higher by 0.3~dex than the non-LTE
corrected values from the D lines. These discrepancies imply that either (1) the
non-LTE corrections applied to these stars are inappropriate, or (2)
measurements of LTE abundances from the very strong D lines and/or very weak
subordinate lines are uncertain. We note, however, that the abundances derived
from the two subordinate lines agree well in CS~22957--027 and HE~0507--1653.
This suggests that either errors exist in the abundance measurements for the D
lines, or in the non-LTE corrections, or in both.

\subsection{Uncertainties}\label{sec:err}

The random errors in abundance measurements obtained from analysis of molecular
bands are estimated by fitting with synthetic spectra calculated by changing the
assumed abundances of carbon and nitrogen, as mentioned in \S~\ref{sec:cn}.
Random abundance errors in the analysis are estimated from the standard error of
the abundances for each species. These values are sometimes unrealistically
small, however, when only a few lines are detected. For this reason, we adopted
the larger of (1) the value for the listed species, or (2) the standard error
derived using the standard deviation of the abundance from individual Fe {\small
I} lines and the number of lines used for the species as estimates of the random
errors. Typical random errors are on the order of 0.1--0.15~dex.

Errors arising from uncertainties in the atmospheric parameters are
evaluated by adopting $\sigma (T_{\rm eff})=100$~K, $\sigma (\log
g)=0.3$~dex, and $\sigma (v_{\rm tur}) =0.3$~km s$^{-1}$ for four
stars, taking the evolutionary status and strength of the molecular
bands into consideration: CS~22174--007, a giant with relatively weak
molecular absorption; CS~22948--027, a giant with strong molecular
absorption; CS~30322--023, a cool giant; and CS~29503--010, a
main-sequence star. The uncertainties estimated for each of our
program stars are taken to be equal to those associated with the one
of these four templates that is closest in evolutionary status and in
strength of the molecular bands. For stars whose effective
temperatures are more uncertain (see \S~\ref{sec:param}) the abundance
uncertainties are estimated assuming $\sigma (T_{\rm eff})=200$~K. For
stars for which the microturbulent velocities are assumed to be
2.0~km~s$^{-1}$, $\sigma (v_{\rm tur}) =0.6$~km s$^{-1}$ is assumed in
obtaining estimates of abundance uncertainties. Finally, we derive the
total uncertainty by adding, in quadrature, the individual errors, and
list them in Table \ref{tab:res}.

The above error estimates do not include possible systematic errors arising from
non-LTE effects. The non-LTE correction for iron abundances is on the order of
+0.2~dex in very metal-poor stars, according to \citet{thevenin99}, although
\citet{gratton99} find no strong effects ($<0.1$~dex)) for dwarfs and red giants
cooler than 5000~K. The effect on the abundance ratios of important elements in
this work (e.g. Ba, Mg) is also 0.1--0.2~dex, depending on the spectral lines
used for the analyses
\citep{asplund05}. 

The effects of inhomogeneity of the stellar atmosphere (the so-called 3D
effects) are also thought to have a significant influence on the determination
of C and N abundances from molecular features \citep{asplund05}. Although
quantitative estimates for 3D effects, coupling with non-LTE ones, are not
available at present, the corrections are possibly on the order of $0.5$~dex or
more (in the negative direction). We do not apply any correction for these
effects in this paper. Instead, we compare below our results with other studies
that derived carbon and nitrogen abundances from molecular bands for
carbon-enhanced and non carbon-enhanced stars.

\subsection{Comparison with Previous Work}\label{sec:comp}

The chemical composition of the extremely metal-poor ([Fe/H]$=-3.1$) star
BS~16929--005 was determined by \citet{honda04b}. Although the wavelength
coverage is different between their study and ours, the results for elements
analyzed in common agree to within 0.15~dex. An exception is Ca, for which our
analysis yields a [Ca/Fe] lower by 0.3~dex. There is no \ion{Ca}{1} line
analyzed in common by the two studies. Our measurement is based on only two
lines, at 4226~{\AA} and 6162~{\AA}. \cite{magain88} and \citet{ryan96} found the
first of these lines yields abundances that are systematically lower than other
Ca~I lines by 0.18~dex. Similar results are found for CS~30312--100, which has
similar stellar parameters to BS~16929--005, and for which the Ca abundance is
determined from 10 \ion{Ca}{1} lines, including the two mentioned above. The
abundances derived from the lines at 4226~{\AA} and 6162~{\AA} are 0.1~dex lower
than the average of the result provided by the other available Ca~I lines. Given
this, and the small number of lines measured by the present analysis (two
lines), the discrepancy of 0.3~dex is possibly a result of random errors.

Two stars from our program with very large carbon enhancements, CS~22948--027
([C/Fe] = +2.1) and CS~29497--034 ([C/Fe] = +2.7), have been studied in detail by
\citet{hill00} and, more recently, by \citet{barbuy05}. Here we compare our
results with those reported by \citet{barbuy05}. The iron and carbon abundances
of CS~22948--027 derived by our present analysis are about 0.2~dex higher than
theirs. This result is well explained by the difference of effective
temperatures adopted in the analyses (our effective temperature is 200~K higher)
. The nitrogen abundance determined by our study is significantly (about
0.7~dex) higher than theirs. This could also, at least in part, be explained by
the difference in the atmospheric parameters adopted by the two studies. Note
that \citet{barbuy05} used CN lines of the red system for the abundance
analysis, while we analyze a band of the CN violet system, which could also
contribute to the discrepancy in the derived N abundances. Our Na abundance is
also 0.35~dex higher than their LTE abundance determination for this element.
The abundances of other elements studied by both works show fairly good
agreement.

For CS~29497--034, the results of our study agree well with those of
\citet{barbuy05}. The exceptions are for Na and Mg, for which we
derive abundances that are 0.4 and 0.5~dex higher for these elements,
respectively. The small differences in the atmospheric parameters adopted by the
two studies cannot explain these discrepancies. The abundances of these two
elements for CS~29497--034, as well as the Na abundance for CS~22948--027, are
determined from only two very strong lines, and small measurement errors in
equivalent widths could result in large abundance errors. We note, however, that
the excesses of Na and Mg in CS~29497--034 are clearly found in the results of
\citet{barbuy05}, as well as in our analysis. 

\subsection{An extended sample of CEMP stars}\label{sec:sample}

For the following discussion we combined the results of recent
abundance studies for CEMP stars \citep{preston01, aoki02a, aoki02b,
aoki02c, aoki02d, depagne02, cohen03, lucatello03, aoki04, cohen04,
barklem05, ivans05, tsangarides05, cohen06, jonsell06} with those of
the present work. A total of 42 stars, including nine for which Na
abundances are determined by our analysis, are compiled. It should be
noted that, although CS~22892--052 is an CEMP star having a large Ba
overabundance, the origin should be quite different from that for
other Ba-enhanced CEMP stars. Hence, this object is not included in
our sample. Details of the selection of the results for objects for
which more than one paper reported chemical abundances are discussed
in the Appendix. The chemical composition and atmospheric parameters
for the extended sample of CEMP stars (64 objects, including 22 stars
studied by the present work) are listed in Table \ref{tab:cempall}.

\section{Carbon Abundance Ratios and a Revised Definition of CEMP stars}\label{sec:cfe}

Figure \ref{fig:cfe} shows the carbon abundance ratios ([C/Fe]) as a function of
iron abundances ([Fe/H]) for our sample (filled circles) and for other CEMP
stars studied by previous work (filled squares). The carbon abundances of these
stars are determined from the strengths of the molecular bands of CH and/or
C$_{2}$. The carbon abundances for other stars that were not regarded as
carbon-enhanced objects by previous studies are indicated with crosses 
\citep{ gratton00, cayrel04, honda04b, aoki05}. The carbon abundances for these
stars were measured from the CH band. We caution that Figure \ref{fig:cfe}
cannot be used to constrain the fraction of CEMP stars as compared to other
metal-poor stars, because of bias in the sample selection. Lucatello et al.
(2006) provides a discussion of these phenomena based on a sample with
better-controlled selection criteria (the HERES sample of Barklem et al. 2005).

Our program stars distribute in metallicity from [Fe/H]$=-1$ to $-3.3$,
providing a useful sample to investigate the metallicity dependence of the
abundance trends of CEMP stars. There are 21 stars in our sample with remarkable
excesses of carbon ([C/Fe] $ > +1$), while the [C/Fe] values of the five other
newly-analyzed stars are not clearly distinguishable from most stars which were
{\it not} identified as carbon-enhanced objects by previous work. Before
proceeding with our inspection of elemental abundance patterns, we first
consider an appropriate definition for the carbon-enhancement phenomenon in
metal-poor stars.

The carbon abundance at the surface of a star is expected to decrease when the
star evolves to become a red giant, due to mixing with internal material that
has been affected by the CNO cycle (the first dredge-up). A further decrease of
carbon, accompanied with an increase of nitrogen, has been observed in highly
evolved metal-poor red giants \citep[e.g.,][]{spite05}, indicating the influence
of extra mixing during evolution along the red-giant branch
\citep{charbonnel95}. As a result, the surface abundance of carbon in late-type
stars will in general be lower than it was when the star was in an earlier
evolutionary stage. These evolutionary effects should thus be included in the
definition of CEMP stars.

For this purpose we calculate the luminosities of the stars in our present
sample, and for other stars studied by previous work, using the relation

$L/L_{\odot}$ ~ $ \propto$ ~ $ (R/R_{\odot})^{2}(T_{\rm eff}/T_{{\rm eff}\odot})^{4}$ ~ $\propto $ ~ $ (M/M_{\odot})(g/g_{\odot})^{-1}(T_{\rm eff}/T_{{\rm eff}\odot})^{4},$

\noindent assuming
the mass of the stars to be 0.8~M$_{\odot}$, following
\citet{aoki05} and \citet{ryan05}. Figure~\ref{fig:cl} depicts [C/Fe]
as a function of luminosity. The [C/Fe] values of most stars studied in previous
works are around +0.4 for $\log(L/$L$_{\odot})\lesssim 2.5$, while the value
decreases with increasing luminosity. A similar trend was obtained by
\citet{gratton00} for their metal-poor field giants (Figure 10 of their paper).
We note that their carbon abundance ratios ([C/Fe]) are systematically lower
than ours by about 0.4~dex, although their abundances were determined from the
CH molecular bands as in our present analyses. This discrepancy might arise
because of the difference in the metallicity ranges studied: the Gratton et al.
sample mostly consists of stars with [Fe/H] $>-2$, while the non carbon-enhanced
stars shown in Figure~\ref{fig:cl} have [Fe/H] $\lesssim -2.5$. The difference
of metallicity could result in differences in the intrinsic carbon abundances,
or in a difference in the 3D effects in stellar atmospheres.
\citet{gratton00} also found decreases in [Li/H], $^{12}$C/$^{13}$C,
and an increase in [N/Fe], at $\log$ ($L/{\rm L}_{\odot}$) $>2$, which are
expected to accompany a decrease of carbon during evolution along the red-giant
branch. We note that some systematic errors possibly exist in our estimates of
luminosity (e.g., NLTE effects on the determination of surface gravities) and
also in carbon abundances (e.g., 3D effects on the molecular bands). However,
such systematic errors do not directly affect the empirical definition of CEMP
stars adopted here.

From inspection of Figure~\ref{fig:cl} we define the stars that satisfy 
the following criteria as CEMP stars:

\begin{enumerate}
\item $[$C/Fe$]\geq +0.7$ for stars with $\log$ ($L/{\rm L}_{\odot}$) $\leq 2.3$

\item $[$C/Fe$]\geq +3.0 - \log(L/$L$_{\odot}$) for stars with $\log
(L/{\rm L}_{\odot}) > $  2.3

\end{enumerate}

\noindent 
The division between CEMP and non-CEMP stars, using the above definition,
is indicated by the dotted line in Figure
\ref{fig:cl}. Four stars with [C/Fe] $< +0.7$ (CS~22174--007,
CS~22886--042, CS~30312--100, and CS~31062--041) are consequently
dropped from the sample of CEMP stars. On the other hand, the cool
giant CS~30322--023 ($\log (L/{\rm L}_{\odot}$)=2.9) is included in the sample of
CEMP stars, although its [C/Fe] value is only +0.56. Another example
of a cool giant with relatively low [C/Fe] that meets our revised
definition of CEMP stars is CS~30314--067 ([C/Fe]=+0.5 and
$\log (L/{\rm L}_{\odot}$)=3.2), which was studied by \citet{aoki02b}. Both cool
giants have quite high nitrogen abundances ([N/Fe] = +2.5 and +1.2 for
CS~30322--023 and CS~30314--067, respectively), suggesting that a
significant amount of carbon was converted to nitrogen during their
red-giant evolution.

With the application of the above criteria, a total of 22 stars in our sample
are classified as CEMP stars. The following discussion is based on these 22
stars and 42 additional stars studied in previous work (shown by filled squares
in Figure~\ref{fig:cl}) that meet our revised definition of CEMP stars. It
should be noted that Figure~\ref{fig:cfe} shows that there is an approximate
limit to the observed [C/Fe] values at a given metallicity. The limit
corresponds to [C/H] $\sim 0$. This point is discussed in more detail in
\S~\ref{sec:ch}.

\section{Discussion}\label{sec:disc}

\subsection{Barium Abundance Ratios}\label{sec:ba}

Barium is a key element for identifying the origin of carbon in CEMP stars (see
\S~\ref{sec:intro}). \citet{ryan05} separated their sample of CEMP
stars into two groups of Ba-enhanced and Ba-normal stars. Figure~\ref{fig:bafe}
shows the barium abundance ratio ([Ba/Fe]) as a function of [Fe/H] for our
expanded sample. A large scatter of barium abundances is known to exist for very
metal-deficient stars \citep[e.g.,][]{mcwilliam98, honda04b}. However, stars
with [Ba/Fe] $ > +0.5$ are quite rare among non carbon-enhanced objects. Here we
define stars with [Ba/Fe] $> +0.5$ as ``Ba-enhanced'' objects, following
\citet{ryan05}\footnote{Note that Beers \& Christlieb (2005) adopted a somewhat
different criterion in their definition of CEMP-no stars ([Ba/Fe] $< 0$) vs.
CEMP-s stars ([Ba/Fe] $> +1.0$; [Ba/Eu] $> +0.5$).}. In Figure \ref{fig:bafe}
the four stars that are excluded from our sample of CEMP stars are shown by open
circles -- the barium abundances of these stars are normal.

One clear result seen from inspection of Figure \ref{fig:bafe} is that 36 of the
37 CEMP stars with [Fe/H] $\geq -2.6$ are Ba-enhanced objects. The exception is
HE~1410+0213 \citep{cohen06}. The iron abundance of this star, determined from
\ion{Fe}{1} lines, is [Fe/H] $= -2.16$. However, note that the [Fe/H] 
derived from \ion{Fe}{2} lines for this object is 0.41~dex lower than that from
\ion{Fe}{1}. Thus, the iron abundance of this star is possibly lower than that
shown in the Figure, and closer to [Fe/H] $= -2.6$. Our first conclusion is that
a majority of, if not all, CEMP stars with relatively high metallicity ([Fe/H]
$\geq -2.6$) also exhibit excesses of the neutron-capture element Ba.

The [Ba/Fe] abundance ratios for the stars in our expanded sample are
shown as a function of [C/Fe] in Figure~\ref{fig:cba} (the upper
panel). A correlation between the barium and carbon abundance ratios
is seen for the Ba-enhanced stars.  The solid line in this Figure
indicates the correlation for stars with [Ba/Fe]$>+0.5$ (54
stars). The null hypothesis that there is no correlation between the
two values is rejected by the regression analysis, as well as by the
Spearman rank correlation test, at the 99.9\% confidence level. This
suggests that the excesses of both the barium and carbon originate in the
same astrophysical site, which is most likely to be AGB progenitor
stars.

As noted above, the carbon abundances of giants are possibly affected
by internal processes, including the CN-cycle. The total abundance of
carbon and nitrogen might thus better represent the initial enrichment
of carbon by progenitor objects. For this purpose, we also show
[Ba/Fe] as a function of [(C+N)/Fe] in the lower panel of
Figure~\ref{fig:cba}. As described above, the CN band is not detected
in seven stars, and only upper limits on nitrogen abundances are
determined. For these stars the upper limit of [(C+N)/Fe] is
determined by adopting the upper limit of the nitrogen abundance,
while the lower limit of [(C+N)/Fe] is determined by assuming the
nitrogen abundance to be zero. The values are listed in
Table~\ref{tab:ciso}.

The correlation between [Ba/Fe] and [(C+N)/Fe] of 44 objects for which
[(C+N)/Fe] values are determined is shown by the dashed line in the
Figure. Although the correlation appears weaker than that found for the [C/Fe]
case, the significance level of the rejection of the no-correlation null
hypothesis is still as high as 99\%. The reason for
the somewhat weaker correlation is the existence of strongly
N-enhanced stars among very metal-poor objects (e.g. CS~22960--053,
CS30322--023). Here we separate the sample into two groups with [Fe/H]
$\geq -2.6$ (29 objects) and [Fe/H] $ < -2.6$ (15 objects). In Figure
\ref{fig:cba}, stars with [Fe/H] $ < -2.6$ are shown by symbols with
open circles, and the correlation is exhibited by the solid line. A
clear correlation between [(C+N)/Fe] and [Ba/Fe] is seen for stars
with higher metallicity, confirming the above suggestion of the
correlation between [Ba/Fe] and [C/Fe]. The slope of the solid line
(0.84 dex per dex) is also interesting. If the [Ba/Fe] and [C/Fe] values are
determined by the enrichment of AGB stars and the dilution in the
secondary objects, a slope of unity is expected. The regression
analysis indicates that this possibility cannot not rejected (at the 70\%
confidence level). A discussion of the stars with lower metallicity can
be found below.

As can be seen in Figure~\ref{fig:bafe}, the 27 stars with [Fe/H] $<
-2.6$ exhibit a large range of Ba abundances ($-1.2 <$ [Ba/Fe] $ <
+3.3$) .\footnote{The three stars with the lowest iron abundances,
  HE~1327--2326 \citep{frebel05}, HE~0107--5240 \citep{christlieb02},
  and G77--61 \citep{plez05}, are depicted in some of our Figures, but
  are not included in the statistics described in this paper.} Nine of
them are Ba-normal stars according to our
definition. Figure~\ref{fig:histfe} shows the [Fe/H] distribution for
Ba-enhanced and Ba-normal CEMP stars, respectively. The Ba-normal
stars are only found in the lowest metallicity range, while
Ba-enhanced stars are distributed over a range in [Fe/H] above
$-3.3$. The null hypothesis that the samples are drawn from the same
parent populations in [Fe/H] is rejected by the rank-sum
(Mann-Whitney) test at the 99.9\% confidence level.  Our second
conclusion is thus that some source(s) of carbon enhancement that are
different from ``standard'' AGB nucleosynthesis must operate
efficiently at low metallicity.


In the smaller sample discussed by \citet{ryan05}, CEMP stars are
clearly distinguished into two groups according to their barium
abundances. However, Figure~\ref{fig:bafe} shows that the CEMP stars
with [Fe/H] $\sim -3$ rather exhibit a continuous distribution in
barium abundances, owing partly to the two CEMP stars having $+0.6 \leq
$ [Ba/Fe] $\leq +1.0$ added by the present work (CS~22960--053 and
CS~30322--023). These two stars have very high nitrogen abundance
ratios, and their values of [Ba/Fe] and [(C+N)/Fe] deviate from the
correlation found for stars with higher metallicity
(Figure~\ref{fig:cba}).  Further observations of stars in this metallicity
range are clearly desirable, in order to derive some more definitive
conclusion on the distribution of Ba abundance ratios at the lowest
metallicity.

\subsection{Distribution of Carbon Abundances}\label{sec:ch}

As discussed in \S~\ref{sec:cfe}, a limit on the carbon excess is seen
in the distribution of carbon abundance ratios
(Figure~\ref{fig:cfe}). This is more clearly seen in Figure
\ref{fig:histc}, where the distributions of carbon abundances ([C/H])
are shown for Ba-enhanced and Ba-normal stars separately.  Most of the
Ba-enhanced stars have $-1 <$ [C/H] $ < 0$, and there is a cutoff of
[C/H] values at around zero (i.e., solar abundance). We note that the
four stars having [C/H] $> 0$ are CS~29503--010, HE~0206--1916,
HE~1447+0102, and CS~22887--048.  The [C/H] values of these stars are,
however, at most +0.14 (see Table~\ref{tab:cempall}), and hence close
to solar. \citet{cohen06} recently stated that the carbon abundances
of their sample of CEMP stars are approximately constant at about
1/5th of the solar value. Most Ba-enhanced CEMP stars in our sample
have carbon abundances that are higher than this value. However, their
sample includes Ba-normal CEMP stars, which could reduce the average
of the carbon abundances. Given this point, and the relatively small
sample size of \citet{cohen06}, the distribution of carbon abundances
derived here is compatible with their conclusion.

The enhancement of carbon in the Ba-enhanced CEMP stars is primarily
determined by the process of AGB nucleosynthesis in the binary
mass-transfer scenario. The material containing enhanced carbon is
transferred to the surface of the companion, and is expected to be
diluted during the evolution of the companion star by mixing with
non-polluted material from the companion's interior. Model
calculations for the nucleosynthesis in AGB stars at low metallicity
have shown that the final predicted yield of carbon is on the order of [C/H] =
0. For instance, \citet{vandenhoek97} provide the total yields of CNO
isotopes in their Table 5.  The $^{12}$C yields in mass are predicted
to be 2.81--4.95$\times 10^{-3}$ M$_{\odot}$, which corresponds to
[C/H] = $-0.02$ -- $+0.23$, for stars with initial mass in the range
1.3--3.0 M$_{\odot}$ and metallicity of $Z=0.001$ ([Fe/H]$ = -1.2$).
\citet{ventura02} derive a similar value ([C/H] = $-0.14$) for a star
with 2.5 M$_{\odot}$ and $Z=0.0002$ ([Fe/H] = $-1.9$). The chemical
compositions of the Ba-enhanced CEMP stars presumably represent the
yields of such low-mass, metal-poor AGB stars. We note that the carbon
yields rapidly decrease with increasing mass ($M\gtrsim
3$M$_{\odot}$), due to hot bottom burning.

Given these model yields of [C/H]~$\sim$~0, the simplest explanation
of the upper cutoff of [C/H]~$\sim$~0 in Figures~\ref{fig:cfe} and
\ref{fig:histc} is that the observed upper envelope 
represents objects in which the bulk of the material originated in the
AGB star and was transferred when the secondary was on or near the
main sequence.  The amount of mass transferred was so large that subsequent
mixing/dilution in the convection envelope of these objects during
giant branch evolution has not greatly changed [C/H] from the
value that existed in material transferred from the AGB donor.




Although the [C/H] values of most Ba-enhanced CEMP stars are found in
a rather narrow range, a correlation between the [C/H] and luminosity
is found. The top panel of Figure \ref{fig:chcnhl} shows [C/H] as a
function of luminosity for the Ba-enhanced CEMP stars (filled symbols)
and the Ba-normal ones (open symbols). A least-square fit to the data
for the Ba-enhanced stars is shown by the solid line.  The null
hypothesis that no correlation exists between these two quantities is
rejected by the regression analysis and by the Spearman rank correlation
test, both at the 99.9\% confidence level. The [C/H] value decreases by
0.6~dex (i.e., a factor of four) with increasing luminosity from $\log
(L/{\rm L}_{\odot})\sim$ 0 to 3. If this correlation reflects the decrease
of carbon abundance at the surface of the object during evolution
along the red giant branch, two mechanisms to decrease carbon might be
considered. One is the dilution of the carbon-rich material provided
by the donor AGB star with the non carbon-rich material of the
secondary we are currently observing. The other is the CN cycle inside
the star through the first dredge-up and extra mixing. However,
the bottom panels of Figure \ref{fig:chcnhl} exhibits a similar
correlation between the [(C+N)/H] values and luminosity to that found for [C/H],
although the confidence level of the null-hypothesis rejection is only 90\% in
this case. That is, the CN cycle is unlikely the primary reason for the
relatively low carbon abundances found in some high luminosity objects. An
exception is CS~30322--023 ([C/H]$=-2.69$ and [(C+N) /Fe]$=-1.46$), whose
surface chemical composition should be severely affected by the CN cycle. We
also shows the plot for [N/H] in the middle panel for completeness, where no
statistically significant correlation is found between the two values.

In summary, some dilution of the envelope material in the receiving
companion is suggested for Ba-enhanced CEMP stars from the correlation
between [C/H] (and [(C+N)/H]) and luminosity.  Most Ba-enhanced CEMP
stars in our sample show, however, that the decrease of [C/H] during
the stellar evolution is less than 1~dex. Given the fact that the
convective layer becomes about two orders of magnitude deeper when a
star evolves to become a red giant, the small dilution of carbon
found in Ba-enhanced CEMP stars suggests that a significant
amount of carbon-enhanced material has been accreted from the donor AGB
star.
 

In contrast to the Ba-enhanced objects, the [C/H] values of Ba-normal
CEMP stars exhibit a very wide distribution, with an average value of
[C/H] that is significantly lower than that of Ba-enhanced stars.  The
difference of the [C/H] distributions between the two groups are
clearly seen in Figure~\ref{fig:histc}. The null hypothesis that the
samples are drawn from the same parent populations in [C/H] is
rejected by the rank-sum test at the 99.9\% confidence level.  This
result suggests again a difference in the mechanisms that enrich
carbon between Ba-normal and Ba-enhanced stars. We note that, since
the average metallicity of Ba-normal CEMP stars is much lower than
that of Ba-enhanced stars (Fig.~\ref{fig:histfe}), their carbon
abundance ratios ([C/Fe]) are higher than the criterion for CEMP stars
(\S 4), although their [C/H] values are not outstandingly high. We
also note that apparent correlations are found between the luminosity
and abundance distributions of C, N, and C+N for Ba-normal stars in
Figure~\ref{fig:chcnhl}. However, they are likely affected by
observational bias. This issue is discussed further in the next subsection. 

\subsection{Evolutionary Status}\label{sec:evo}

\citet{ryan05} suggested that a difference exists in the evolutionary
stages between the Ba-enhanced and Ba-normal CEMP stars. In their
sample, the Ba-normal CEMP stars are found only at the top of the
giant branch, while Ba-enhanced stars are found throughout the H-R
diagram. The excess of unevolved Ba-enhanced CEMP stars, relative to
Ba-normal ones, was attributed by \citet{ryan05} to the dilution of
surface carbon enhancements in the Ba-enhanced stars during the
evolution from the main-sequence to the red giant branch.

To examine this hypothesis for our expanded sample of CEMP stars,
Figure~\ref{fig:histl} shows the luminosity distribution of
Ba-enhanced and Ba-normal CEMP stars, respectively (see also Table
\ref{tab:cempall}). The luminosity, as estimated from our adopted
atmospheric parameters as mentioned in \S \ref{sec:cfe}, is an
indicator of the stars' present evolutionary status. The Ba-enhanced
stars distribute over a wide range of luminosity. We note that the
number of objects at $\log (L/{\rm L}_{\odot}) =1$--1.5 is relatively
small.  This probably corresponds to the gap between giants and
subgiant/main-sequence stars. Although no Ba-normal CEMP main-sequence
star was included in our previous sample \citep{ryan05}, two such
stars were reported by \citet{cohen04} and \citet{cohen06}.

The suggested difference in the luminosity distributions between the
Ba-enhanced and Ba-normal CEMP stars is not clear for the present
sample. Indeed, according to the rank-sum  test the
difference of the luminosity distributions between these two groups
shown in Figure~\ref{fig:histl} is not statistically significant (the
null hypothesis of identical luminosity distributions for the two
samples can be rejected at only the 46\% significance level). Thus,
the existence of different luminosity distributions for the
Ba-enhanced and Ba-normal CEMP stars suggested by \citet{ryan05} is
not confirmed by our expanded sample. Moreover, as discussed in
\S~\ref{sec:ch}, the carbon abundance distribution of Ba-enhanced CEMP
stars (Figure~\ref{fig:histc}) shows a concentration around [C/H] $ >
-1$, regardless of evolutionary status, indicating that the dilution
of the transferred material with carbon enhancement is not as significant
as \citet{ryan05} presumed.


It should be kept in mind that the luminosity distribution of observed
CEMP stars could be affected by a selection bias, because the
strengths of the molecular bands are dependent on the stellar
temperature, and, therefore, on the luminosity of the
star. Figure~\ref{fig:chteff} shows the [C/H] values as a function of
effective temperature. In this Figure the Ba-normal and Ba-enhanced
stars are indicated by open and filled symbols, respectively. All
Ba-enhanced stars but one (CS~30322--023) have [C/H] $\gtrsim -1$,
while most of Ba-normal stars have lower carbon abundances. The CH
bands that are used in the usual abundance studies are easily detected
even in main-sequence stars if the carbon abundance is as high as
solar (i.e., [C/H] $\sim 0$). However, if the abundance is lower than
[C/H] $\sim -1.5$, quite high-quality spectra are required for
detection of the CH band. In order to estimate the detection limit of
the CH G band, we calculate synthetic spectra for model atmospheres of
dwarfs ($T_{\rm eff}$=6000--7000~K and $\log g$=4.0), subgiants
($T_{\rm eff}$=5500~K and $\log g$=3.5) and giants ($T_{\rm
eff}$=4000--5000~K and $\log g$=1.0--2.5). The dotted line in Figure
\ref{fig:chteff} indicates the [C/H] values required for the depth of
the CH G band to be about 2~\%, which we take as a conservative limit
for detection of this feature in high-quality high-resolution
spectra. The detection limit of the CH G band in medium-resolution
spectra, by which CEMP stars are usually initially identified in
surveys, is higher than the values shown by this line\footnote{This is
not the case for non-CEMP stars, for which no selection from the
G-band is made in the sample selection.}. Even if Ba-normal CEMP
main-sequence stars having similar carbon excesses to those found in
Ba-normal CEMP giants exist, they would not be identified as CEMP
stars in the surveys. On the other hand, carbon-enhanced giants are
detected even if the carbon enhancement is not as significant as those
of CEMP main-sequence stars.  Ba-enhanced dwarfs as well as giants are
detected because of their very high carbon excesses. This could lead
to differences in the observed luminosity distributions of the two
types of CEMP stars.

\subsection{Other Elements}\label{sec:others}

Figure~\ref{fig:aball} shows the abundance ratios ([X/Fe]) as
functions of [Fe/H] for elements other than C, N, and Ba. The
abundances of non-CEMP stars are adopted from \citet{cayrel04},
\citet{honda04b}, and \citet{aoki05}. Besides these sources, the
results of \citet{gratton00} are also shown for Na abundances, while
those of \citet{stephens02} are shown for Na, Mg, Ca, Ti, Cr, and Ni
abundances. The distributions of the abundance ratios of Sc, Ti, Cr,
Ni, and Zn in CEMP stars are quite similar to those in non
carbon-enhanced stars. A few stars appear to deviate from the observed
trends. For instance, HE~1319--1935 ([Fe/H] $ = -1.8$) has [Cr/Fe] $=
+0.75$, and CS~30338--089 has [Sc/Fe] $=1.26$ ([Fe/H] $=-2.45$) (both
stars are Ba-enhanced objects). However, these results are based on
measurements for only one line, and no clear interpretations for these
stars can be made at present. 

Several CEMP stars exhibit quite high Ca abundances compared to the
typical value of non-CEMP stars ([Ca/Fe] $\sim +0.3$). In our sample,
three stars (HE~1429--0551, HE~1528--0409, and HE~2221--0453) have
[Ca/Fe] $> +0.8$. However, the abundances are determined from only 1--3
\ion{Ca}{1} lines in 5500--6200~{\AA}. The three objects are
relatively cool and exhibit large carbon excesses; numerous
C$_{2}$ and CN molecular lines exist in this wavelength range. Hence,
the Ca abundances of such objects are possibly overestimated due to
contamination by molecular features. Five stars from the literature have
[Ca/Fe] $ > +0.8$ (Table~\ref{tab:cempall}). One is CS~31062--050, studied
by \citet{aoki02c}, who measured the abundance from two \ion{Ca}{1}
lines in the blue range. However, these lines are possibly affected by
CH molecular lines. Indeed, an analysis for 15 \ion{Ca}{1} lines in
the red spectrum of CS~31062--050, which is used for the analysis of
Na lines in the present work, results in [Ca/Fe] = +0.40 (contamination
by C$_{2}$ and CN molecular lines is not severe in the red spectrum
of this object). We conclude that our previous measurement
overestimated the Ca abundance for this object. The other four
objects having [Ca/Fe] $ > +0.8$ were studied by \citet{cohen06}, who
measured the Ca abundances for blue lines as \citet{aoki02c} did. The four
stars are cool giants having large excesses of carbon. Although we cannot
further investigate the Ca abundances of these stars, because red
spectra for them are not available, our re-analysis for CS~31062--050 suggests
that the analyses for blue lines in CEMP stars possibly overestimate
the Ca abundances. Hence, we here do not regard the large Ca
overabundances found in several objects as general characteristics of CEMP
stars.

In contrast, a large range of abundance ratios is found for Na and
Mg. We find moderate overabundances of Mg for most of the CEMP stars
in our sample, as is usually observed for non carbon-enhanced stars,
but three stars in our sample exhibit large excesses of Mg. One such
star is HE~1447+0102 ([Fe/H] $= -2.5$), whose Mg abundance ([Mg/Fe] $=
+1.43$) is determined from the Mg {\small I} 5172~{\AA} line alone,
with a correspondingly large uncertainty. The Mg abundances of
CS~29497--034 ([Mg/Fe] $= +1.31$, [Fe/H] $= -2.91$) and CS~29528--028
([Mg/Fe] $= +1.69$, [Fe/H] $= -2.86$) are, however, based on
measurements for three lines, and the scatter of abundances derived
from individual lines is not significantly large. Therefore, we
suggest that these two stars have distinguishable excesses of Mg from
other metal-poor stars. It should be noted that \citet{hill00} and
\citet{barbuy05} reported an excess of this element in CS~29497--034,
although the Mg abundance derived by their work is not as high as our
present result (see \S~\ref{sec:comp}). Other stars with excesses of
Mg studied by previous work are LP~625--44 ([Fe/H] $= -2.7$; Aoki et
al. 2002a), CS~22949--037 ([Fe/H] $= -4.0$; e.g., Depagne et
al. 2002), and CS~29498--043 ([Fe/H] $= -3.5$; e.g., Aoki et
al. 2004). The origin of the Mg excesses in the latter two stars has
been suggested to be ``faint supernovae'' explosions that ejected only
a small amount of material surrounding the iron core
\citep{tsujimoto03,umeda03}. On the other hand, the sources of Mg in
other Mg-enhanced CEMP stars are still unknown.
 We note that LP~625--44, CS~22949--037, and
CS~29498--043 also exhibit large overabundances of oxygen (see
references above). For the two stars with Mg excesses in our sample,
no useful constraint on the oxygen abundance has been obtained at
present. Measurements of oxygen abundances for these stars, possibly
from the triplet lines at 7770~{\AA} or near-infrared CO lines, are
strongly recommended.

The Na abundances of CEMP stars exhibit a large range of
enhancements. Although some uncertainties arising from non-LTE
corrections and the treatment of damping exist in the determinations
of Na abundances, this large scatter cannot be fully explained by
measurement errors alone. Previous evidence \citep{pilachowski96,
gratton00} points to there being a slight increase in [Na/Fe] in more
evolved, metal-poor red giants. The former work suggests values around
[Na/Fe] = $-0.25$ for field subgiants, [Na/Fe] =$ -0.17$ for field
giants, and possibly higher values in globular-cluster giants. Many
Ba-enhanced CEMP stars deviate in their sodium abundances from the
solar ratios. In order to investigate the correlation between the
excesses of carbon and Na, we show [Na/Fe] as a function of [(C+N)
/Fe] in Figure \ref{fig:cnna}. In this Figure, Ba-enhanced and
Ba-normal CEMP stars are shown by filled and open symbols,
respectively. A clear correlation is found for the 14 Ba-enhanced
stars in our sample (filled circles) for which C, N, and Na abundances
are determined (as confirmed by the rank correlation test at the 99\%
confidence level), while the correlation is weaker for the full sample
of Ba-enhanced objects. We can at least conclude that Na-enhanced
stars also have high [(C+N)/Fe] values. The Mg-enhanced stars
CS~29497--034 and CS~29528--028 exhibit very large enhancements of Na
([Na/Fe] = +2.74 and +1.78, respectively). The iron abundances of
these two stars are very low ([Fe/H] $\sim -2.9$) while their excesses
of C, N, and Ba are very high, as compared to the rest of our
sample. These two stars can be regarded as extreme cases of the
Ba-enhanced CEMP stars.

Sodium abundances have also been determined for seven Ba-normal CEMP
stars in the present sample (in addition to these stars, HE~0107--5240
and HE~1327--2326, the objects having [Fe/H]$<-5$, and the dwarf
carbon star G77-61, are also shown by open triangles in the
Figure). Four stars among them do not deviate significantly from solar
ratios, or have lower values. The carbon excess in these stars is
moderate ([C/Fe] $\sim +1.0$). This class of objects was first
reported by \citet{aoki02b}, and possibly is a separate class from the
other three Ba-normal, extremely carbon-enhanced objects. The two
Ba-normal $\alpha$-element-enhanced CEMP stars, CS~22949--037 and
CS~29498--043, both have large enhancements of Na.  The other
extremely carbon-rich CEMP star, CS~22957--027, also exhibits a large
excess of Na. Although the abundances of $\alpha$-elements in
CS~22957--027 are normal, this star could be related in some manner to
the above two $\alpha$-element-enhanced objects.

%
%

\subsection{Origins of Carbon Excess in CEMP stars}

As described in \S~\ref{sec:intro}, the fraction of carbon-enhanced
objects is significant at low metallicity, although the estimate of this
fraction is dependent on the definition of carbon-enhanced stars and
the methods used to identify them. For instance, \citet{cohen05} have
claimed that the fraction of CEMP stars in a sample of giants with
[Fe/H] $< -2.0$ is 14.5 $\pm 4$\%. Lucatello et al. (2006),
based on high-resolution analyses of a much larger sample of stars, including
many earlier-type stars than the Cohen et al. sample, conclude that this
fraction is at least 20 $\pm 2$\%. An important question is what is (are) the
origin(s) of the carbon-enhancement in such objects.  


Our results indicate that the majority of CEMP stars are Ba-enhanced
objects -- the fraction of Ba-enhanced objects is more than 80\% in
our sample. This trend is particularly remarkable for stars with
relatively high metallicity, above [Fe/H] $= -2.6$. Therefore, our
conclusion is that the large fraction of CEMP stars at low metallicity
can be principally attributed to the high efficiency of AGB
nucleosynthesis followed by the mass transfer to a surviving companion
\citep{komiya06}.

At the lowest metallicities, the contribution of Ba-normal stars to
the population of CEMP stars becomes large. This is another factor
that leads to the increased fraction of carbon-enhanced stars 
at low metallicity.  The most extreme cases are the
carbon-enhanced hyper metal-poor stars with [Fe/H] $ < -5$
HE~1327--2326 \citep{frebel05} and HE~0107--5240 \citep{christlieb02},
for which only upper limits of Ba are determined.

Given the wide distribution of carbon abundances among CEMP stars, as
well as the unique abundance patterns of several stars in this class,
the CEMP population of stars almost certainly comprises several
astrophysical origins. No well-established model exists to explain the
Ba-normal CEMP stars, although \citet{ryan05} suggested that they form
from carbon-rich gas (from a massive progenitor-star population) in
contrast to the AGB progenitor paradigm, which explains many features
of Ba-enhanced objects.

For the two stars CS~22949--037 and CS~29498--043, large excesses of
the $\alpha$-elements and oxygen are also found, suggesting the
contribution of ``faint supernova'' nucleosynthesis, with the
progenitor stars undergoing mixing and fallback \citep{tsujimoto03,
umeda03}. A similar interpretation might be applicable to the hyper
metal-poor stars with [Fe/H] $< -5.0$ (see also Iwamoto et al. 2005),
although several other models have been proposed for these stars.  One
idea is that such stars might have formed from gas polluted by mass
loss from early-generation massive, rapidly-rotating stars with [Fe/H]
$< -6.0$ \citep{hirschi06, meynet06}. This model has the additional
advantage of being able to account for the otherwise mysterious lack
of detected Li in the turnoff star HE~1327--2326
\citep{hirschi06b}. Piau et al. (2006) have argued that this same
population might even be responsible for the alteration of primordial
Li from (higher) Big Bang Nucleosynthesis values down to a level
consistent with the Spite Plateau Li abundance found in metal-poor
turnoff stars. If a significant population of such primordial objects
did indeed exist, one might argue that they could be closely related
to the astrophysical origin of the Ba-normal CEMP stars found at quite
low metallicity. Karlsson (2006) has appealed to the same progenitors
to account for the possible gap in the metallicity distribution of
halo stars between [Fe/H] $\sim -4.0$ and $\sim -5.0$ revealed by the
HK and HES prism surveys. Also of interest is the recent paper by
Chiappini et al. (2006), where their new chemical evolution models
{\it require} the large yields of carbon and nitrogen from the models
discussed by \citet{hirschi06} in order to account for the upturn of
the observed [N/O] and [C/O] ratios in the ``unmixed'' carbon-{\it
normal} stars described by Spite et al. (2005).

Other possibilities for the origin of Ba-normal CEMP stars, in
particular for the objects having normal abundance ratios of the
$\alpha$-elements, exist.  \citet{cohen06} suggested that one possible
interpretation could be that the s-process in AGB stars with such low
metallicities did not produce any distinctive excess of
barium. \citet{komiya06} have recently proposed that the Ba-normal
CEMP stars are the result of nucleosynthesis in extremely metal-poor
AGB stars with masses higher than 3.5~M$_{\odot}$. Their models of AGB
stars, originally developed by \citet{fujimoto00}, suggest that the
s-process does not occur in extremely metal-poor AGB stars in this
mass range because of the low efficiency of radiative $^{13}$C burning
(the case IV' in their classification).  Interestingly, according to
their models, this class of AGB stars should have metallicities of
[Fe/H] $< -2.5$, which is very similar to the metallicity range over
which Ba-normal CEMP stars are found (\S \ref{sec:ba}). Unfortunately,
there is no evidence of binarity for most objects in this
class. However, according to their models, binary systems having very
long orbital periods are able to produce carbon-enhancement by stellar
winds from a primary AGB star at very low metallicity. For such
long-period binary systems, the present absence of variations in
radial velocity is not conclusive. A statistical analysis (e.g.,
Lucatello et al. 2005) based on large samples of Ba-normal CEMP stars
observed over very long temporal baselines will be required in order
to test such models.

\section{Concluding Remarks}

In this paper we have determined chemical abundances for 26 metal-poor
stars, and classified 22 stars among them as Carbon-Enhanced
Metal-Poor (CEMP) stars, based on a new definition for the
carbon-enhancement phenomenon including the effects of evolution along
the giant branch on observed carbon abundances.  Sodium abundances
were determined for another nine CEMP stars for which abundances of
other elements were previously studied. Combining the results of
previous work for 42 CEMP stars and the 22 other CEMP stars in our
present sample, we investigated the abundance trends found in CEMP
stars and discussed the origins of carbon excesses in these
objects. We found that most CEMP stars (more than 80\%) exhibit
excesses of neutron-capture elements (e.g., Ba), suggesting
significant contributions of AGB nucleosynthesis. However, the
fraction of Ba-normal stars increases at lower metallicity. The
difference in the metallicity distribution of Ba-enhanced and
Ba-normal CEMP stars found in the present work should be a key to
understanding the processes responsible for the Ba-normal CEMP stars.

Another difference between the Ba-enhanced and Ba-normal stars is
found in their carbon abundance distributions. While the Ba-enhanced
stars have quite high carbon abundances ($-1 \lesssim $ [C/H] $ \lesssim
0$), independent of their metallicity, the [C/H] values of Ba-normal
CEMP stars distribute over a wide range, while their average value is
relatively low. This difference also suggests that the processes
responsible for the Ba-normal CEMP stars are quite different from the
AGB nucleosynthesis that can explain the chemical abundances of
Ba-enhanced CEMP stars.


Some possible interpretations are proposed for the origin of Ba-normal
CEMP stars, in particular for the $\alpha$-element-enhanced
stars. Moreover, these stars are likely to be closely related to the
two known hyper metal-poor stars with [Fe/H] below $-5$. However, the
sample of such stars studied with high-resolution spectroscopy is
still quite small (ten objects in the present sample).  Further
observations for this class of objects are strongly recommended.

Measurements of abundances for Ba-enhanced CEMP stars based on
high-resolution spectroscopy have been rapidly increasing in recent
years. Although the abundances of these objects are usually
interpreted within the paradigm of AGB nucleosynthesis and mass
transfer in binary systems, their detailed abundance patterns are not
well explained by current models. One example is the abundance pattern
found for objects with very large excesses of neutron-capture
elements, which cannot be explained by the present AGB models. Several
CEMP stars are known to have enhancements of both r- and s-process
elements, the so-called CEMP-r/s stars. More detailed abundance
studies, as well as the monitoring of radial velocities to investigate
their binarity, are strongly desired. We plan to investigate these
problems in future papers in this series.

\acknowledgments

W.A. is grateful for useful discussions on the nucleosynthesis and
evolution of CEMP stars with Drs. M. Y. Fujimoto and T. Suda.
J.E.N. acknowledges support from Australian Research Council grant
DP0342613. T.C.B. acknowledges partial support from a series of grants
awarded by the US National Science Foundation, most recently, AST
04-06784, as well as from grant PHY 02-16783; Physics Frontier
Center/Joint Institute for Nuclear Astrophysics
(JINA). N.C. acknowledges support from Deutsche Forschungsgemeinschaft
under grants Ch~214/3 and Re~353/44.

\appendix

\section{CEMP Stars Investigated in the Present Work}

In this paper we discussed the abundances and evolutionary status for
22 CEMP stars analyzed by the present work and 42 CEMP stars that have
been previously studied. Table~\ref{tab:cempall} lists the abundances
and atmospheric parameters of CEMP stars discussed in \S~\ref{sec:cfe}
and \S~\ref{sec:disc}. References are provided in the last column of
the Table. The chemical compositions of the following objects were
reported by more than one paper. Here we provide a few comments for
these objects and justify our adopted results.

\begin{itemize}

\item {\bf CS~22948--027 and CS~29497--034}: These objects were
studied in considerable detail by \citet{hill00} and
\citet{barbuy05}. Comparisons of their measurements and ours are
discussed in \S~\ref{sec:comp}. \citet{preston01} studied binarity, as
well as abundances, for CS~22948--027. In the discussion we adopted
the results from the present work.

\item {\bf CS~22957--027}: \citet{norris97b} and \citet{bonifacio98}
reported abundances for this star. \citet{preston01} studied its
binarity, as well as its elemental abundances. We adopted the results
of our previous work \citep{aoki02c, aoki02d}, based on spectra of
higher quality than previously available. The Na abundance of this
object was determined by the present work. \citet{preston01} also
reported a large excess of Na, although their value of [Na/Fe] is
0.5~dex lower than ours. We note that our result is based on
abundances from subordinate lines and the non-LTE corrected abundances
from D lines (see \S \ref{sec:naba}).

\item {\bf CS~22898--027}: \citet{mcwilliam95} reported the abundances
of this object for the first time. Its binarity and abundances were
studied by \citet{preston01}. We adopted the results of our previous
work \citep{aoki02c, aoki02d}, based on spectra with remarkably higher
quality than available previously.  The Na abundance was determined in
the present work.

\item {\bf CS~22942--019}: We adopted the results of our previous
studies \citep{aoki02c, aoki02d}. This star was also analyzed by
\citet{preston01}, who derived results similar to ours.  We adopted
their Na abundance.

\item {\bf CS~31062--050}: The abundances determined by our previous
work \citep{aoki02c, aoki02d} were adopted. This object was also
studied by \citet{johnson04}, based on high quality spectra. Their
results for elements discussed in the present paper agree within the
errors with our previous work. An exception is the Ba abundance ratio,
for which \citet{johnson04} derived a value 0.5~dex higher than ours.

\item {\bf CS~22880-074}: The results of our previous studies
\citep{aoki02c, aoki02d} and the present work for the Na abundance
were adopted. This object was also studied by \citet{preston01}. The
abundances derived by them agree within 0.3~dex with our results. An
exception is the Mg abundance, for which \citet{preston01} derived a
quite low value ([Mg/Fe] = 0.06) compared to other metal-poor stars.

\item {\bf CS~29502--092}: The abundances determined by
\citet{aoki02a} were adopted. This object was more recently studied by
\citet{tsangarides05}, who compared in some detail their results with
those of \citet{aoki02a}.

\item {\bf CS~29497--030}: This object was studied by \citet{sneden03}
and \citet{sivarani04}. More recently, \citet{ivans05} made a more
comprehensive abundance analysis using higher quality spectra. We
adopted the results of \citet{ivans05}.

\item {\bf CS~22949--037}: The abundances of this object were reported
by \citet{mcwilliam95} for the first time. This object was
re-investigated by \citet{norris01,norris02}, \citet{depagne02}, and
\citet{cayrel04}. We adopted the results by \citet{depagne02}, who
made the most comprehensive abundance analysis. An exception is the Na
abundance, for which we adopted the result of \citet{cayrel04}, who
corrected the value of \citet{depagne02} using the same spectra.

\item {\bf CS~22183--015}: This object was studied by
\citet{johnson02}, in particular for the neutron-capture
elements. Abundances of this object, including those of light elements
not measured by \citet{johnson02}, were recently reported by
\citet{cohen06}. Here we adopted the results by \citet{cohen06}.

\end{itemize}





\clearpage

\begin{deluxetable}{llccrrrrr}
\tabletypesize{\small}
\tablewidth{0pt}
\tablecaption{OBJECT LIST AND OBSERVING LOG \label{tab:obs}}
\tablehead{
ID & Object & RA(2000) & Dec(2000) & Exp. Time & Count & Obs.date (UT) & $V_{\rm r}$ & s.e.\tablenotemark{a} \\
   &        &          &                 & (sec) & (5200~{\AA}) &      & km~s$^{-1}$ &  km~s$^{-1}$  \\
}
\startdata
1 & BS~16929--005 & 13:03:29.4 & +33:51:06 & 600 & 1800 & May 25, 2003 & $-$50.4 & 0.13 \\
2 & CS~22174--007 & 01:14:06.7 & $-$11:02:32 & 600 & 6700 & Oct 26, 2002 & $-$70.76 & 0.03 \\
3 & CS~22886--042 & 22:20:25.9 & $-$10:23:20 & 600 & 2300 & Oct 26, 2002 & $-$221.4 & 0.05 \\
4 & CS~22948--027 & 21:37:45.4 & $-$39:27:20 & 600 & 3100 & Oct 26, 2002 & $-$62.49 & 0.09 \\
5 & CS~22960--053 & 22:16:17.0 & $-$43:54:21 & 2400 & 1900 & Oct 26, 2002 & 282.26 & 0.08 \\
6 & CS~29503--010 & 00:04:55.4 & $-$24:24:19 & 900 & 2500 & Oct 26, 2002 & $-$24.28 & 0.04 \\
7 & CS~29528--028 & 02:28:26.5 & $-$21:33:00 & 1200 & 1500 & Oct 26, 2002 & \nodata  & \nodata \\
  & CS~29528--028 & \nodata    & \nodata     & 3000 & 3000 & Oct 28, 2002 & 58.23 & 0.34 \\
\enddata
\tablenotetext{a}{Standard error of radial velosity $\sigma / \sqrt{n}$, 
where $\sigma$ is the standard deviation of the values from
individual lines and $n$ is the number of lines used.}
\tablenotetext{b}{CS~29497--034 = HE~0039--2635}
\end{deluxetable}

\begin{deluxetable}{llcccccccccr}
\tabletypesize{\footnotesize}
\rotate
\tablewidth{0pt}
\tablecaption{PHOTOMETRIC DATA \label{tab:photo}}
\tablehead{
ID & Object & $B$ & $V$ & $R$ & $I$ & $J$ & $H$ & $K$ & $E(B-V)$ & $T_{\rm eff}(V-K)$ & $\Delta T_{\rm eff}$\tablenotemark{a} \\
}
\startdata
1 & BS~16929--005 &  14.228 &  13.610 &  13.208 &  12.766 &  12.170 &  11.750 &  11.670 &    0.01 & 5257 & 31 \\
2 & CS~22174--007 &  13.103 &  12.409 &  11.961 &  11.509 &  10.885 &  10.404 &  10.318 &    0.08 & 5309 & $-$21 \\
3 & CS~22886--042 &  14.106 &  13.275 &  12.763 &  12.269 &  11.511 &  11.028 &  10.940 &    0.03 & 4830 & $-$36 \\
4 & CS~22948--027 &  13.790 &  12.660 &  12.160 &  11.760 &  10.979 &  10.534 &  10.427 &    0.08 & 5135 & $-$298 \\
5 & CS~22960--053 &  15.705 &  14.950 &  14.537 &  14.046 &  13.298 &  12.833 &  12.761 &    0.07 & 5160 & $-$121 \\
6 & CS~29503--010 &  14.209 &  13.737 &  13.430 &  13.128 &  12.726 &  12.468 &  12.433 &    0.08 & 6532 & $-$33 \\
7 & CS~29528--028 &  14.862 &  14.510 &  14.292 &  14.015 &  13.755 &  13.535 &  13.537 &    0.03 & 7046 & 248 \\
8 & CS~30312--100 &  13.629 &  12.916 &  12.444 &  11.995 &  11.390 &  10.946 &  10.879 &    0.05 & 5256 & 105 \\
\enddata
\tablenotetext{a}{$\Delta(T_{\rm eff}) = T_{\rm eff}(V-K)_{0} - <T_{\rm eff}(V-R)_{0}, T_{\rm eff}(V-I)_{0}, T_{\rm eff}(R-I)_{0}>$}
\end{deluxetable} 

\begin{deluxetable}{llccrrrcc}
\tabletypesize{\small}
\tablewidth{0pt}
\tablecaption{OBJECT LIST AND OBSERVING LOG \label{tab:naobs}}
\tablehead{
ID & Object & Exp. Time & Count        & Obs.date (UT)  \\
           & (sec)     & (5900~{\AA}) &                \\
}
\startdata
Na1 & HD~196944      &  30 &17500 & Sep 13, 2005  \\
Na2 & LP~706--7      & 180 & 2500 & Sep 18, 2005  \\
Na3 & CS~22183--015\tablenotemark{a}  & 600 & 2200 & June 17, 2005  \\
Na4 & CS~22880--074  &3000 & 5500 & June 18, 2005  \\
Na5 & CS~22898--027  &3000 & 15000 & July 3, 2000  \\
Na6 & CS~22957--027  &5330 & 17800 & June 18, 2005  \\
Na7 & CS~29526--110  & 720 & 3500 & Sep 18, 2005  \\
Na8 & CS~30301--015  & 300 & 1900 & Sep 13, 2005  \\
Na9 & CS~31062--050  &3600 &22200 & June 29, 2005  \\
\tableline
\enddata
\tablenotetext{a}{CS~22183--015 = HE~0058--0244}
\end{deluxetable}

\begin{deluxetable}{lcccccccccccc}
\tabletypesize{\footnotesize}
\tablewidth{0pt}
\tablecaption{EQUIVALENT WIDTHS \label{tab:ew1}}
\tablehead{
Species & Wavelength & L.E.P. & $\log gf$ & \multicolumn{9}{c}{$W_{\lambda}$ (m{\AA})} \\
        &   ({\AA})  &  (eV)  &           & (1) & (2) & (3) & (4) & (5) & (6) & (7) & (8) & (9) \\
}
\startdata
     Na I &   5688.20&    2.10&   $-0.46$&   ... &   ... &   ... &   ... &   ... &   5.3  &   38.8&   ... &   11.1\\
     Na I &   5889.95&    0.00&   $ 0.10$&   95.5&  161.9&  164.8&  195.8&   ... &  153.7 &  194.6&  147.8&   ... \\
     Na I &   5895.92&    0.00&   $-0.20$&   64.2&  141.5&  137.6&  188.5&  172.0x&  147.2 &  183.8&  126.0&   ... \\
     Mg I &   4571.10&    0.00&   $-5.69$&   ... &   54.6&   58.1&   ... &   ... &   26.9&   ... &   38.5&   99.5\\
     Mg I &   4702.99&    4.35&   $-0.52$&   28.6&   67.1&   70.5&   ... &   ... &   ... &   55.8&   63.2&   ... \\
     Mg I &   5172.69&    2.71&   $-0.38$&  128.5&  191.9&  174.9&  191.1&  130.4&   ... &  141.4&  180.3&  190.5\\
     Mg I &   5183.60&    2.72&   $-0.16$&  143.1&  215.5&  190.2&   ... &  146.1&   ... &  157.3&  202.4&   ... \\
     Mg I &   5528.40&    4.35&   $-0.49$&   28.0&   69.0&   66.9&   ... &   41.8&  103.6&   ... &   64.7&   71.0\\
\enddata
\end{deluxetable}

\begin{deluxetable}{lcccccccccccc}
\tabletypesize{\footnotesize}
\tablewidth{0pt}
\tablecaption{EQUIVALENT WIDTHS \label{tab:ew2}}
\tablehead{
Species & Wavelength & L.E.P. & $\log gf$ & \multicolumn{9}{c}{$W_{\lambda}$ (m{\AA})} \\
        &   ({\AA})  &  (eV)  &           & (10) & (11) & (12) & (13) & (14) & (15) & (16) & (17) & (18) \\
}
\startdata
     Na I &   5682.63&    2.10&   $-$0.70&   ... &   ... &   ... &   ... &   ... &   ... &   36.0&   ... &   ... \\
     Na I &   5688.20&    2.10&   $-$0.46&   14.9&   ... &   ... &   ... &   34.2&   ... &   60.6&   ... &   ... \\
     Na I &   5889.95&    0.00&    0.10&  156.6&  166.8&  218.6&  200.6&  225.6&  186.9&  295.2&  154.9&  202.6\\
     Na I &   5895.92&    0.00&   $-$0.20&  151.3&  151.6&  202.3&  178.0&  216.0&  169.0&  246.3&  143.2&  183.4\\
     Mg I &   4571.10&    0.00&   $-$5.69&   37.0&   64.3&   ... &   ... &   80.1&   55.9&  103.4&   ... &   ... \\
     Mg I &   4702.99&    4.35&   $-$0.52&   ... &   73.3&   ... &   ... &   ... &   46.0&   ... &   ... &   ... \\
     Mg I &   5172.69&    2.71&   $-$0.38&  177.0&  180.1&  177.6&  175.3&   ... &  173.2&   ... &  133.1&  189.0\\
     Mg I &   5183.60&    2.72&   $-$0.16&   ... &   ... &  229.4&  236.9&   ... &   ... &   ... &  144.5&   ... \\
\enddata
\end{deluxetable}

\begin{deluxetable}{lccccccccccc}
\tabletypesize{\footnotesize}
\tablewidth{0pt}
\tablecaption{EQUIVALENT WIDTHS \label{tab:ew3}}
\tablehead{
Species & Wavelength & L.E.P. & $\log gf$ & \multicolumn{8}{c}{$W_{\lambda}$ (m{\AA})} \\
        &   ({\AA})  &  (eV)  &           & (19) & (20) & (21) & (22) & (23) & (24) & (25) & (26)  \\
}
\startdata
     Na I &   5889.95&    0.00&    0.10&   ... &  228.5&  171.7&   ... &  170.8&   ... &  292.6&  166.6 \\
     Na I &   5895.92&    0.00&   $-$0.20&  246.6&  211.5&  149.7&  213.9&  162.2&  246.5&  275.9&  142.8 \\
     Mg I &   4571.10&    0.00&   $-$5.69&   ... &   ... &   ... &   ... &   ... &   ... &   ... &   33.0 \\
     Mg I &   4702.99&    4.35&   $-$0.52&   ... &   ... &   ... &   ... &   ... &   ... &   ... &   67.0 \\
     Mg I &   5172.69&    2.71&   $-$0.38&  297.7&  187.9&  192.1&  216.7&  170.4&  272.2&  183.4&  163.8 \\
     Mg I &   5183.60&    2.72&   $-$0.16&   ... &   ... &   ... &   ... &   ... &   ... &   ... &  182.5 \\
     Mg I &   5528.40&    4.35&   $-$0.49&   ... &   ... &   ... &   ... &   ... &   ... &  114.0&   53.5 \\
     Ca I &   5588.76&    2.53&    0.36&  119.2&   ... &   ... &   80.1&   ... &   ... &   52.3&   31.8 \\
\enddata
\end{deluxetable}

\begin{deluxetable}{lcccc}
\tabletypesize{\footnotesize}
\tablewidth{0pt}
\tablecaption{EQUIVALENT WIDTHS \label{tab:ew4}}
\tablehead{
Species & Wavelength & L.E.P. & $\log gf$ & $W_{\lambda}$ (m{\AA}) \\
        &   ({\AA})  &  (eV)  &           & CS~22958--028  \\
}
\startdata
     Fe I &   3618.77&    0.99&    0.00&      68.4 \\      
     Fe I &   3631.46&    0.96&   $-$0.04&      47.5  \\     
     Fe I &   3745.56&    0.09&   $-$0.77&      65.0   \\    
     Fe I &   3758.23&    0.96&   $-$0.03&      52.2   \\    
     Fe I &   3815.84&    1.49&    0.23&      65.1  \\     
     Fe I &   3820.43&    0.86&    0.12&      92.4 \\      
     Fe I &   3824.44&    0.00&   $-$1.36&      44.0  \\     
     Fe I &   3825.88&    0.92&   $-$0.04&      76.5  \\
\enddata
\end{deluxetable}

\begin{deluxetable}{lcccccccccccc}
\tabletypesize{\footnotesize}
\tablewidth{0pt}
\tablecaption{EQUIVALENT WIDTHS OF Na LINES \label{tab:ew5}}
\tablehead{
Species & Wavelength & L.E.P. & $\log gf$ & \multicolumn{9}{c}{$W_{\lambda}$ (m{\AA})} \\
        &   ({\AA})  &  (eV)  &           & (Na1)& (Na2)& (Na3)& (Na4) & (Na5)& (Na6)& (Na7) & (Na8) & (Na9) \\
}
\startdata
     Na I &   5682.63&    2.10&   $-$0.70&   ... &   ... &   ... &   ... &   ... &   3.4 &   ... &   ...  & ... \\
     Na I &   5688.20&    2.10&   $-$0.46&   ... &   ... &   ... &   ... &   ... &   8.6 &   ... &   ...  & 6.1 \\
     Na I &   5889.95&    0.00&    0.10&  186.7&  146.3&  116.6&  151.4& 128.9 &  171.3&   84.1&  220.3 & ... \\
     Na I &   5895.92&    0.00&   $-$0.20&  165.2&  122.3&  100.8&  133.8&  99.9 &  155.0&   ... &  197.8 & ... \\
\tableline
\enddata
\end{deluxetable}

\begin{deluxetable}{llcccc}
\tablewidth{0pt}
\tablecaption{ATMOSPHERIC PARAMETERS \label{tab:param}}
\tablehead{
ID & Object & $T_{\rm eff}$ & $\log g$ & [Fe/H] & $v_{\rm micr}$ \\
   &        & (K)         &            &        & (km$^{-1}$) \\
}
\startdata
 1 & BS~16929--005 & 5250 & 2.8 & $-$3.2 & 2.3 \\
 2 & CS~22174--007 & 5300 & 2.7 & $-$2.3 & 2.1 \\
 3 & CS~22886--042 & 4800 & 1.6 & $-$2.5 & 2.1 \\
 4 & CS~22948--027 & 5000 & 1.9 & $-$2.3 & 2.0 \\
 5 & CS~22960--053 & 5200 & 2.1 & $-$3.2 & 2.1 \\
 6 & CS~29503--010 & 6500 & 4.5 & $-$1.2 & 1.9 \\
 7 & CS~29528--028 & 6800 & 4.0 & $-$2.9 & 1.9 \\
 8 & CS~30312--100 & 5300 & 2.8 & $-$2.4 & 2.2 \\
 9 & CS~30322--023 & 4300 & 1.0 & $-$3.3 & 3.3 \\
10 & CS~30338--089 & 5000 & 2.1 & $-$2.5 & 1.7 \\
11 & CS~31062--041 & 5100 & 2.2 & $-$2.5 & 2.3 \\
12 & CS~29497--034 & 4900 & 1.5 & $-$3.0 & 2.2 \\
13 & HE~0206--1916 & 5200 & 2.7 & $-$2.2 & 2.0 \\
14 & HE~0400--2030 & 5600 & 3.5 & $-$1.8 & 2.2 \\
15 & HE~0441--0652 & 4900 & 1.4 & $-$2.5 & 2.3 \\
16 & HE~0507--1653 & 5000 & 2.4 & $-$1.5 & 2.0$^{a}$ \\
17 & HE~1005--1439 & 5000 & 1.9 & $-$3.2 & 2.0 \\
18 & HE~1157--0518 & 4900 & 2.0 & $-$2.4 & 2.0$^{a}$ \\
19 & HE~1319--1935 & 4600 & 1.1 & $-$1.8 & 2.0$^{a}$ \\
20 & HE~1429--0551 & 4700 & 1.5 & $-$2.5 & 2.0 \\
21 & HE~1447+0102 & 5100 & 1.7 & $-$2.5 & 1.8 \\
22 & HE~1523--1155 & 4800 & 1.6 & $-$2.2 & 1.8 \\
23 & HE~1528--0409 & 5000 & 1.8 & $-$2.6 & 1.8 \\
24 & HE~2221--0453 & 4400 & 0.4 & $-$2.2 & 2.0$^{a}$ \\
25 & HE~2228--0706 & 5100 & 2.6 & $-$2.5 & 1.6 \\
26 & HE~2330--0555 & 4900 & 1.7 & $-$2.8 & 1.8 \\
\tableline
\enddata
\tablenotetext{a}{assumed values (see text)}
\end{deluxetable}


\clearpage
\pagestyle{empty}
\begin{deluxetable}{lcccccccccccccc}
\rotate
\tabletypesize{\footnotesize}
\tablewidth{0pt}
\tablecaption{ABUNDANCE RESULTS \label{tab:res}}
\tablehead{
   & Fe (Fe I)\tablenotemark{a} &  Fe (Fe II)\tablenotemark{a} & C\tablenotemark{b} & N\tablenotemark{b} & Na & Mg & Ca & Sc & Ti (Ti I) & Ti (Ti II) & Cr  & Ni  & Zn  & Ba \\
}
\startdata
$\log \epsilon_{\rm Sun}$\tablenotemark{c} & 7.45 & 7.45 & 8.39 & 7.78 & 6.17 & 7.53 & 6.31 & 3.05 & 4.90 & 4.90 & 5.64 & 6.23 & 4.60 & 2.17 \\
\hline
BS~16920--005   &  &  &  &  &  &  &  &  &  &  &  &  &  &  \\
\hline
$\log \epsilon$         &  4.28 & 4.27 & 6.3 & $<7.0$ & 2.94 & 4.85 & 3.27 & ... & ... & 2.17 & 2.14 & ... & ... & $-$1.48 \\
{[}X/Fe]         & $-$3.17 & $-$3.18 & 1.08 & $<$2.39 & $-$0.06 & 0.49 & 0.13 & ... & ... & 0.44 & $-$0.34 & ... & ... & $-$0.48 \\
N              &   24  & 3 & CH4322 & CN4215 & 2 & 4 & 2 & ... & ... & 8 & 1 & ... & ... & 2 \\
s.e.([X/Fe])&  0.14 & 0.13 & 0.16 & 0.07 & 0.17 & 0.10 & 0.09 & ... & ... & 0.14 & 0.09 & ... & ... & 0.12 \\
\hline
CS~22174--007   &  &  &  &  &  &  &  &  &  &  &  &  &  &    \\
\hline
$\log \epsilon$ & 5.26 & 5.26 & 6.6 & $<7.0$ & 4.36 & 5.82 & 4.35 & 0.96 & 3.09 & 3.17 & 3.26 & 4.07 & 2.46 & $-$0.59 \\
\enddata
\tablenotetext{a}{[Fe/H] values are given in the lines for [X/Fe].}
\tablenotetext{b}{Molecular bands used for analyses are given in the line for the number of lines used (``$N$''). CH4322: the CH $\lambda$4322 band; CC5170: the C$_{2}$ $\lambda$5167; CC5635: the $\lambda$5635 band; CN4215: the CN $\lambda$4215 band.}
\tablenotetext{b}{Solar elemental abundances derived by \citet{asplund05b}.}
\end{deluxetable}
\clearpage
\pagestyle{plaintop}

\begin{deluxetable}{lccccc}
\tablewidth{0pt}
\tablecaption{LUMINOSITY AND ABUNDANCE RATIOS OF CARBON ISOTOPES AND NITROGEN \label{tab:ciso}}
\tablehead{
Object & $\log L/({\rm L}_{\odot})$ & [(C+N)/Fe] & $^{12}$C/$^{13}$C & $^{12}$C/N & $^{13}$C/N \\
}
\startdata
BS~16920--005 &  1.46 &  0.98--1.76 &  ...  &  $>0.20$ &  ...\\
CS~22174--007 &  1.57 &  0.31--0.86 &  ...  &  $>0.40$ &  ...\\
CS~22886--042 &  2.50 &  0.02--1.35 &  ...  &  $>0.05$ &  ...\\
CS~22948--027 &  2.27 &  2.20 & $10\pm 3$& 2.00  &  0.20\\
CS~22960--053 &  1.87 &  2.50 &  ...  &  0.40 &  ...\\
CS~29503--010 &  0.13 &  0.97--1.12 &  ...  &  $>2.51$ &  ...\\
CS~29528--028 &  0.71 &  2.67--3.09 &  ...  &  $>0.63$ &  ...\\
CS~30312--100 &  1.47 &  0.44--1.22 &  ...  &  $>0.20$ &  ...\\
CS~30322--023 &  2.91 &  1.79 &  ...  &  0.05 &  ...\\
CS~30338--089 &  2.07 &  1.98 & $12\pm 4$ &  25.12 &   2.09\\
CS~31062--041 &  2.01 &  0.48--1.03 &  ...  &  $>0.40$ &  ... \\
CS~29497--034 &  2.64 &  2.70 & $12\pm 4$ & 5.01  &  0.42 \\
HE~0206--1916 &  1.54 &  2.04 & $15\pm 5$ &  12.59 &   0.84\\
HE~0400--2030 &  1.14 &  2.09 &  ...  &   0.10 &  ...\\
HE~0441--0652 &  2.74 &  1.32 &  ...  &  12.59 &  ...\\
HE~0507--1653 &  1.77 &  1.23 & $40^{+20}_{-10}$ & 12.59  &  0.31\\
HE~1005--1439 &  2.27 &  2.41 &  ...  &  19.95 &  ...\\
HE~1157--0518 &  2.14 &  2.08 & $15\pm 5$  & 15.85  &  1.06\\
HE~1319--1935 &  2.93 &  1.37 & $8\pm 3$  &  39.81 &   4.98\\
HE~1429--0551 &  2.56 &  2.20 & $30^{+20}_{-10}$  & 31.62 &   1.05 \\
HE~1447+0102 &  2.51 &  2.39 & $25\pm 10$  & 50.12   & 2.00\\
HE~1523--1155 &  2.50 &  1.83 &  ...  &  6.31 &  ...\\
HE~1528--0409 &  2.37 &  2.37 & $12\pm 5$ &  10.00 &   0.83\\
\tablebreak
HE~2221--0453 &  3.55 &  1.75 & $10\pm 4$  &  39.81 &   3.98\\
HE~2228--0706 &  1.61&   2.23 & $15^{+5}_{-3}$  & 63.10  &  4.21\\
HE~2330--0555 &  2.44 &  2.00 &  ...  &50.12  & ... \\
\tableline
\enddata
\end{deluxetable}

\begin{deluxetable}{lcccc}
\tablewidth{0pt}
\tablecaption{ABUNDANCE RESULTS FOR NEUTRON-CAPTURE ELEMENTS in CS~29528--028 \label{tab:heavy}}
\tablehead{
Element & $\log \epsilon$ & [X/Fe] & $N$ & s.e. \\
}
\startdata
Sr      &    1.87         &   1.81 &  2 &  0.20     \\
Zr      &    1.98         &   2.25 &  1 &  0.20     \\
La      &    1.20         &   2.93 &  6 &  0.20    \\
Ce      &    1.63         &   2.91 &  1 &  0.20     \\
Nd      &    1.54         &   2.95 &  1 &  0.20    \\
\tableline
\enddata
\end{deluxetable}

\begin{deluxetable}{llcccc}
\tablewidth{0pt}
\tablecaption{NA ABUNDANCES AND NLTE CORRECTIONS \label{tab:nares}}
\tablehead{
ID & Objects & $\log \epsilon$(Na)$_{\rm LTE}$ & $\Delta_{\rm NLTE}$ &  $\log \epsilon$(Na)$_{\rm NLTE}$ & [Na/Fe]$_{\rm NLTE}$ \\
}
\startdata
1 & BS~16929--005 & 2.94 & $-$0.3 & 2.64 & $-$0.36 \\
2 & CS~22174--007 & 4.27 & $-$0.4 & 3.87 & $-$0.10 \\
3 & CS~22886--042 & 3.91 & $-$0.4 & 3.51 & $-$0.05 \\
4 & CS~22948--027 & 4.88 & $-$0.4 & 4.48 & 0.52 \\
5 & CS~22960--053 & ... & ... & ... & ... \\
6 & CS~29503--010 & 5.27 & ... & 5.27 & 0.16 \\
7 & CS~29528--028 & 5.99 & $-$0.7 & 5.64 & 2.33 \\
8 & CS~30312--100  & 4.02 & $-$0.4 & 3.62 & $-$0.23 \\
\enddata
\end{deluxetable}

\begin{deluxetable}{lrrrrrrrrrrrrrrrl}
\tabletypesize{\tiny} {\footnotesize}
\rotate
\tablewidth{0pt}
\tablecaption{ELEMENTAL ABUNDANCES OF CEMP STARS \label{tab:cempall}}
\tablehead{
Object & [Fe/H] & [C/Fe] & [N/Fe] & [Na/Fe] & [Mg/Fe]  & [Ca/Fe] & [Sc/Fe] & [Ti/Fe] & [Cr/Fe] & [Ni/Fe] & [Zn/Fe] & [Ba/Fe] & $T_{\rm eff}$ & $\log g$ & $\log L/L_{\odot}$ & Ref.\tablenotemark{a} \\
}
\startdata
BS~16920--005 & $-$3.17&  1.08&  $<$2.39& $-$0.36&  0.49&  0.13& \nodata& 0.44 & $-$0.34& \nodata& \nodata& $-$0.48& 5250&  2.80&  1.46 &       \\
CS~22948--027 & $-$2.21&  2.12&  2.43&  0.52&  0.55&  0.57& \nodata&  0.64& $-$0.12& \nodata&  0.31&  2.31& 5000&  1.90&  2.27  &       \\
CS~22960--053 & $-$3.14&  2.05&  3.06& \nodata&  0.65&  0.43&  0.49& 0.54 & $-$0.23& \nodata& \nodata &  0.86& 5200&  2.10&  2.14  &       \\
CS~29503--010 & $-$1.06&  1.07&  $<$1.28&  0.16&  0.36&  0.16&  0.31&  0.39& $-$0.12&  0.01&  0.32&  1.50& 6500&  4.50&  0.13  &       \\
CS~29528--028 & $-$2.86&  2.77&  $<$3.58&  2.33&  1.69&  0.46&  0.59& 0.87 & \nodata&  0.26& \nodata&  3.27& 6800&  4.00&  0.71  &       \\
CS~30322--023 & $-$3.25&  0.56&  2.47&  1.04&  0.54&  0.30&  0.28&  0.21& $-$0.36& \nodata& \nodata&  0.59& 4300&  1.00&  2.91  &       \\
CS~30338--089 & $-$2.45&  2.06&  1.27&  0.46&  0.48&  0.51&  1.26&  0.55& $-$0.01& \nodata&  0.46&  2.22& 5000&  2.10&  2.07  &       \\
HE~0039--2635 & $-$2.91&  2.72&  2.63&  1.37&  1.31&  0.52& \nodata&  0.46& $-$0.15& \nodata& \nodata&  2.23& 4900&  1.50&  2.64  &       \\
\enddata
\tablenotetext{a}{References --- (1) \citet{aoki04}; (2) \citet{aoki02c}; (3)
\citet{aoki02d}; (4) \citet{aoki02b}; (5) \citet{aoki02a}; (6)
\citet{ivans05}; (7) \citet{preston01}; (8) \citet{cohen03}; (9)
\citet{lucatello03}; (10) \citet{depagne02}; (11) \citet{cohen04};
(12) \citet{jonsell06}; (13) \citet{tsangarides05}; (14)
\citet{barklem05}; (15) \citet{cohen06}}
\tablenotetext{a}{[Ca/Fe]=0.40 is derived by the present work (see \S~\ref{sec:others}).}
\end{deluxetable}

\clearpage

\begin{figure} 
\begin{center}
\includegraphics[width=8.5cm]{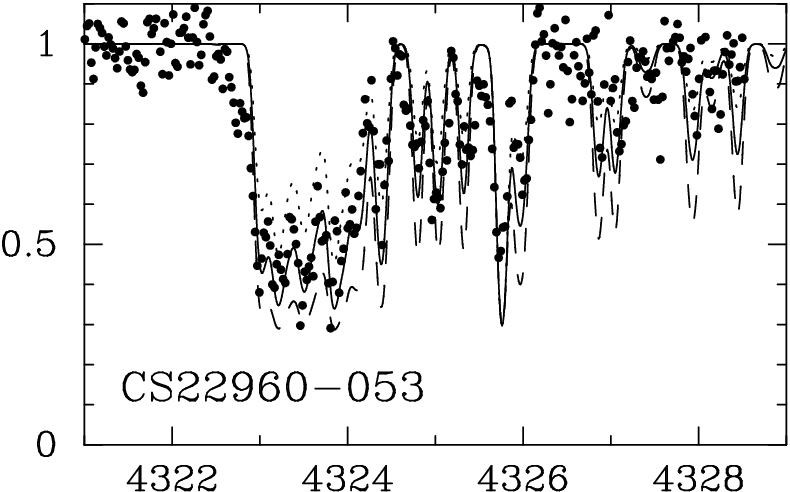} 
\includegraphics[width=8.5cm]{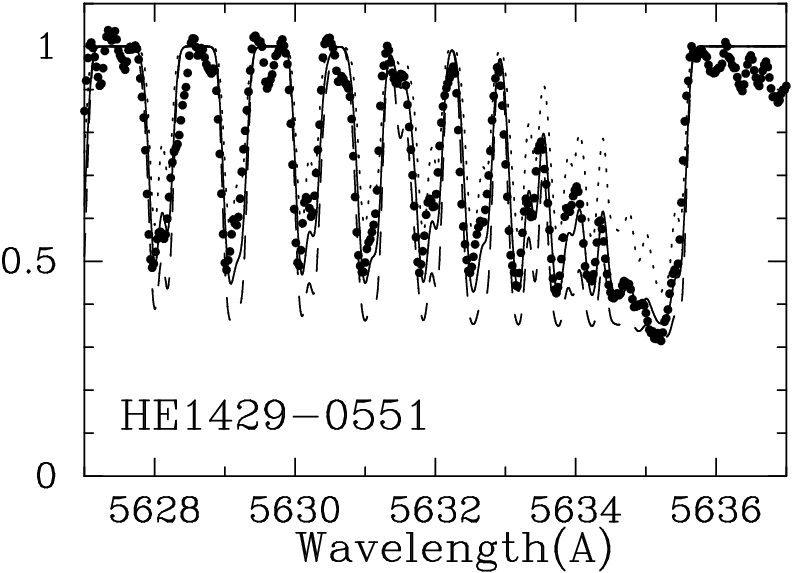} 
\caption[]{Examples of the CH and C$_{2}$ bands used for carbon abundance
  determinations for CS~22960--053 and HE~1429--0551,
  respectively. The solid lines indicate the synthetic spectra for
  adopted carbon abundances, while other lines show those for carbon
  abundances changed by $\pm$0.3~dex and $\pm$0.2~dex for
  CS~22960--053 and HE~1429--0551, respectively. 
}\label{fig:chc2}
\end{center}
\end{figure} 
 
\begin{figure} 
\includegraphics[width=8.5cm]{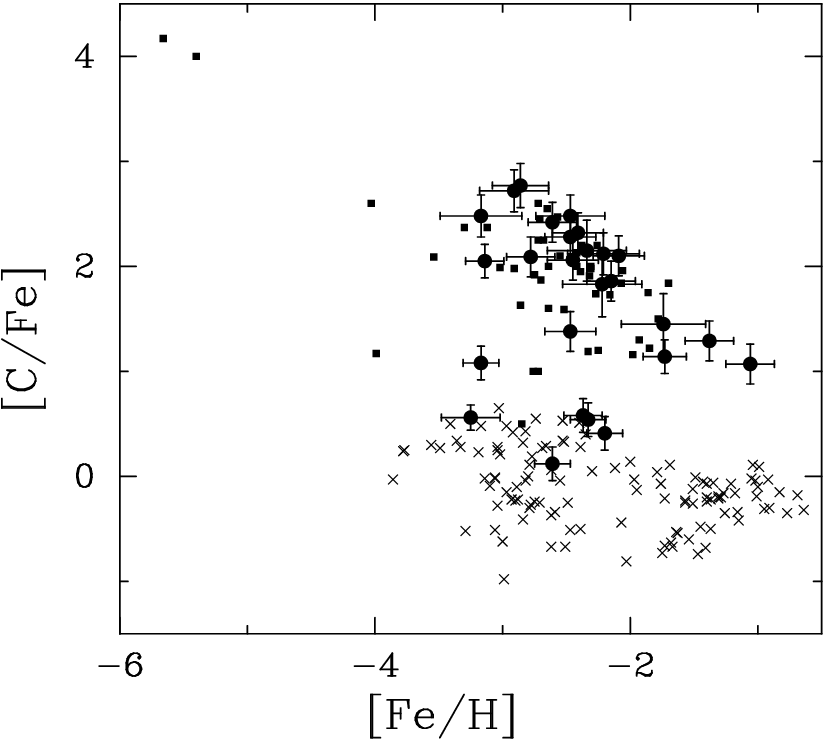} 
\caption[]{The carbon abundance ratio ([C/Fe]) as a function of
  [Fe/H]. The filled circles (with error bars) indicate stars from our
  sample, while filled squares indicate other CEMP stars. Non
  carbon-enhanced objects studied by previous works \citep{gratton00,
  cayrel04, honda04b, aoki05} are shown by crosses.}
\label{fig:cfe}
\end{figure}

\begin{figure} 
\includegraphics[width=8.5cm]{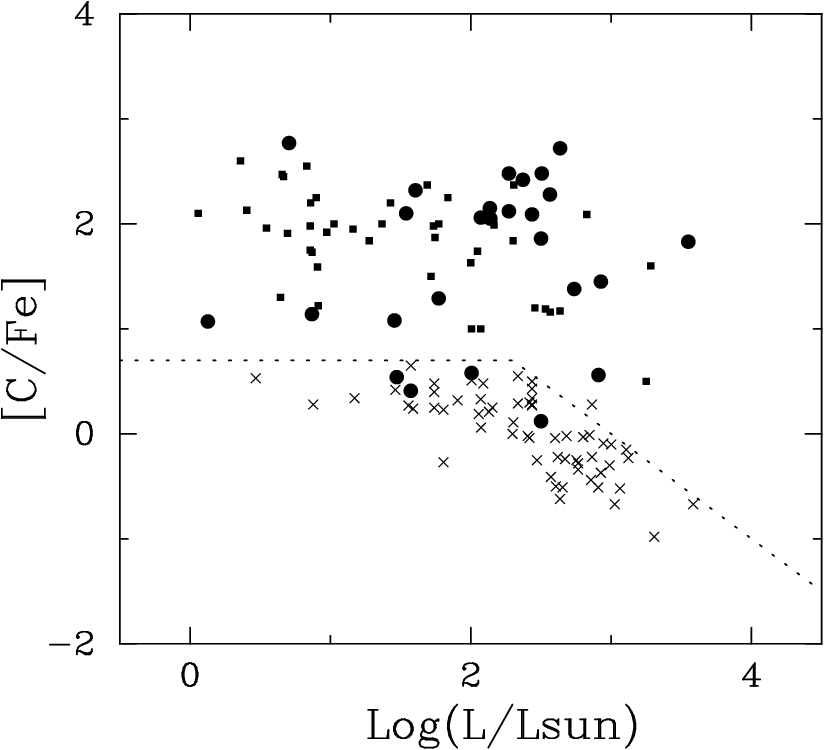} 
\caption[]{The carbon abundance ratio ([C/Fe]) as a function of
luminosity estimated from measured effective temperatures and
gravities for stars in our expanded sample. The meanings of symbols
are the same as in Figure~\ref{fig:cfe}. The results for non
carbon-enhanced stars from \citet{cayrel04}  \citet{honda04b} and \citet{aoki05} are
shown here. The dashed line indicates the dividing line between
stars that we define as CEMP and the carbon-normal metal-poor 
stars (see text) }\label{fig:cl}
\end{figure} 

\begin{figure} 
\includegraphics[width=8.5cm]{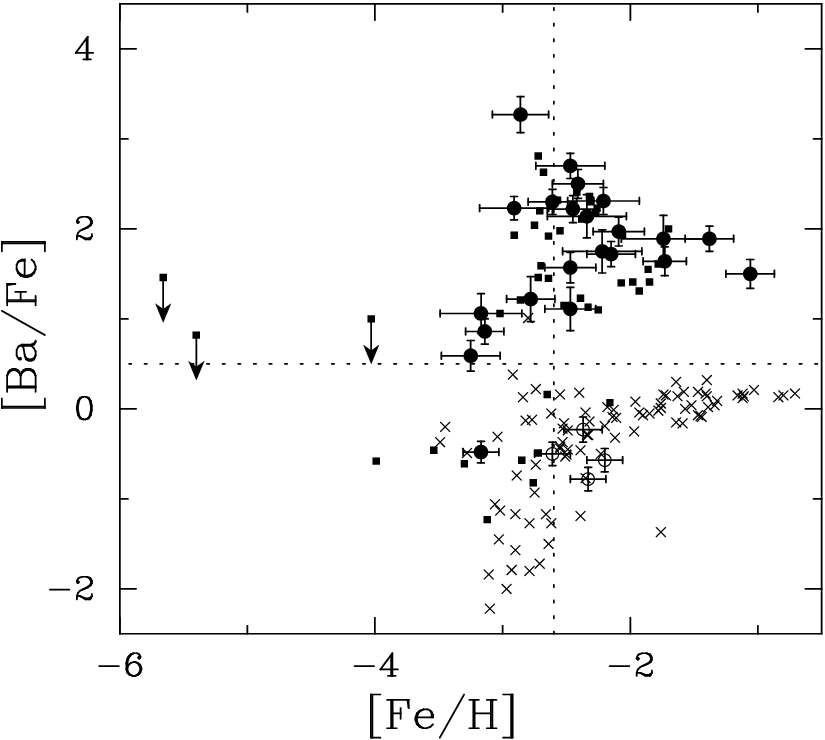} 
\caption[]{ The same as Figure \ref{fig:cfe}, but for [Ba/Fe].  The
results for non carbon-enhanced stars from \citet{stephens02},
\citet{cayrel04} \citet{honda04b} and \citet{aoki05} are shown
here. The non carbon-enhanced objects in our sample (see text) are
shown by open circles here. The upper limits of [Ba/Fe] are shown for
HE~1327--2326, HE~0107--5240, and G~77--61. The dotted lines indicate
[Ba/Fe]$=+0.5$ and [Fe/H]$= -2.6$. }\label{fig:bafe}
\end{figure} 

\begin{figure} 
\begin{center}
\includegraphics[width=8.5cm]{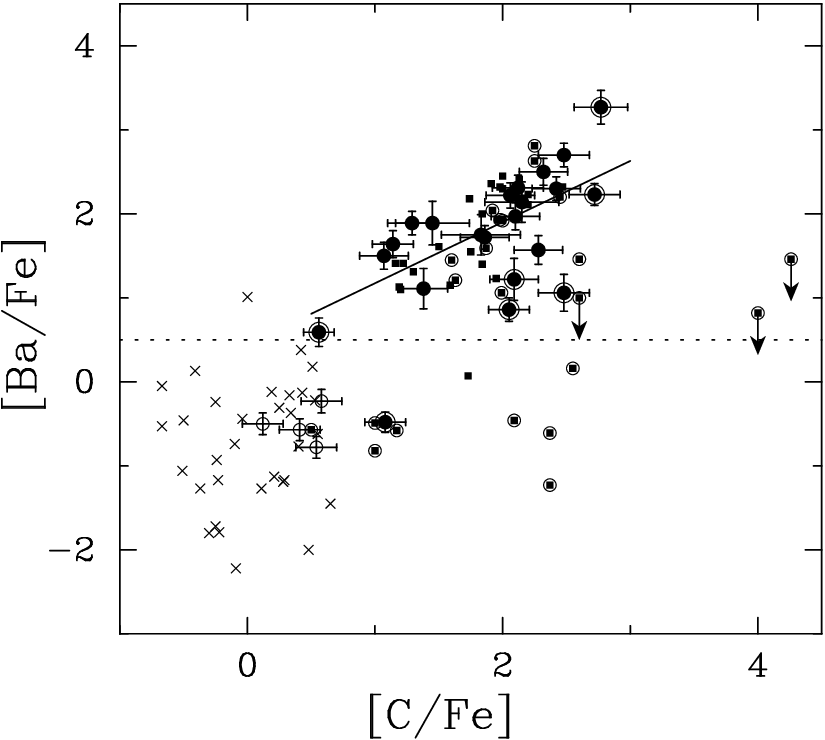} 
\includegraphics[width=8.5cm]{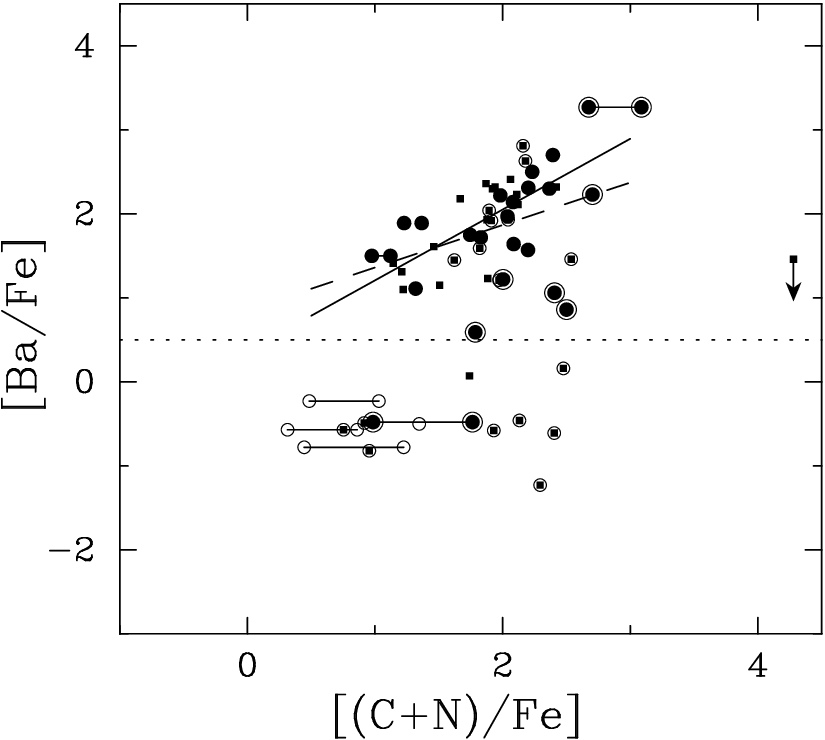} 
\caption[]{The [Ba/Fe] ratio as a function of [C/Fe] for stars in our
expanded sample (upper panel). The meanings of the symbols are the
same as in Figure~\ref{fig:bafe}. Stars with [Fe/H] $< -2.6$ are
indicated by symbols with large open circles. The solid line indicates
the least square fit for stars with [Ba/Fe]$>+0.5$ (54 stars).  The lower
panel is the same as the upper one, but for [Ba/Fe] as a function of
[(C+N)/Fe] instead of [C/Fe]. The upper and lower limits of [(C+N)/Fe]
are connected by lines for stars for which only upper limits to the N
abundance are determined (see text). The dashed line indicates the
least square fit for stars with [Ba/Fe]$>+0.5$ for which N abundance is
determined (44 stars), while the solid line shows that for stars
having [Fe/H]$\geq -2.6$ among them (29 stars).}
\label{fig:cba}
\end{center}
\end{figure} 

\begin{figure} 
\includegraphics[width=8.5cm]{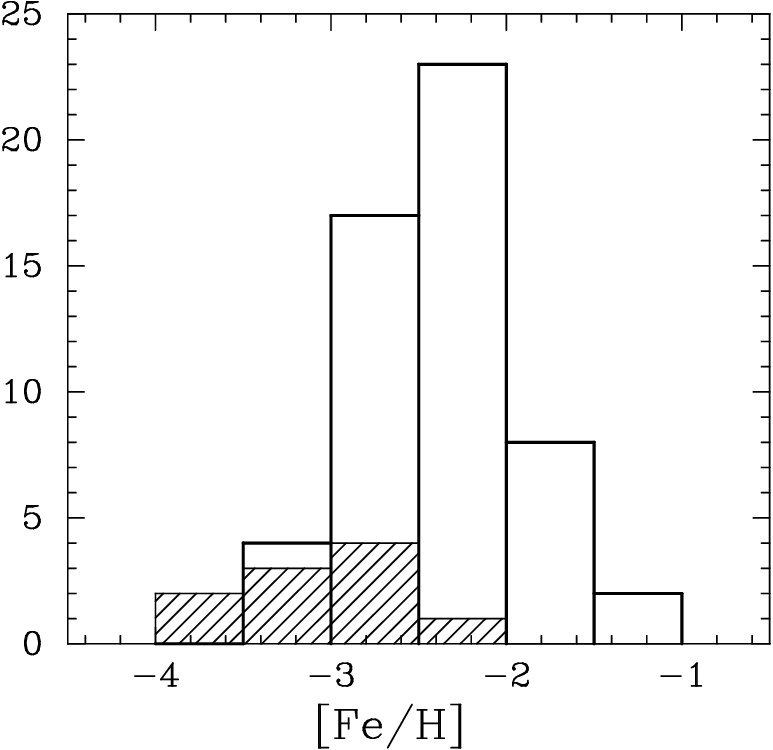} 

\caption[]{ The [Fe/H] distribution for CEMP stars in our expanded
  sample. The histogram with the ``open'' bars indicates the Ba-enhanced
  stars, while that with the hatched area is for the Ba-normal stars.
  There appears to be a clear difference in the distributions of
  [Fe/H] for these two classes of stars (see
  text).}\label{fig:histfe}.

\end{figure} 

\begin{figure} 
\includegraphics[width=8.5cm]{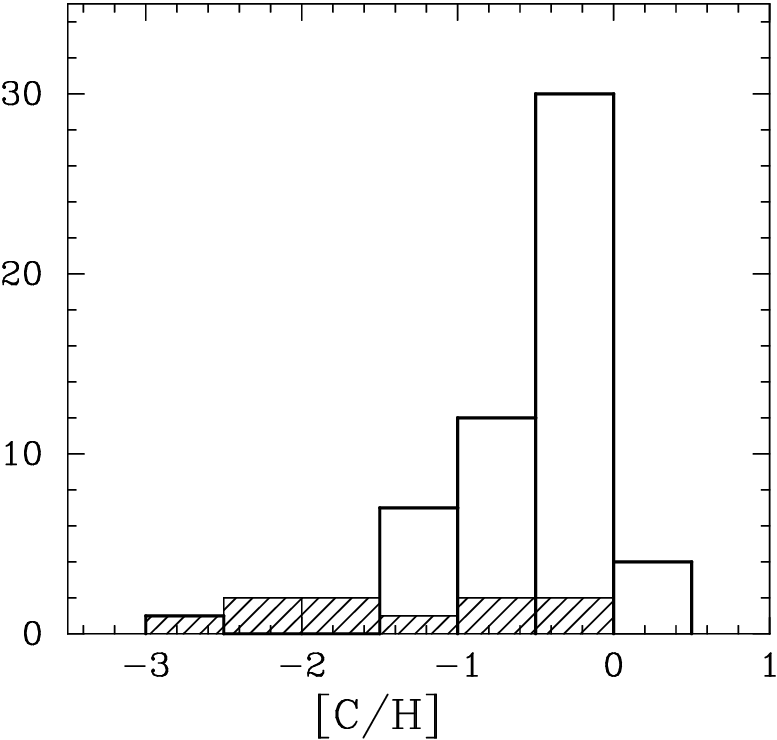} 
\caption[]{ The same as Figure~\ref{fig:histfe}, but for the distribution of [C/H]
for the CEMP stars in our expanded sample. Again, there appears a clear
difference in the distribution of [C/H] for the Ba-enhanced and Ba-normal
stars (see text).}
\label{fig:histc}
\end{figure} 

\begin{figure} 
\includegraphics[width=8.5cm]{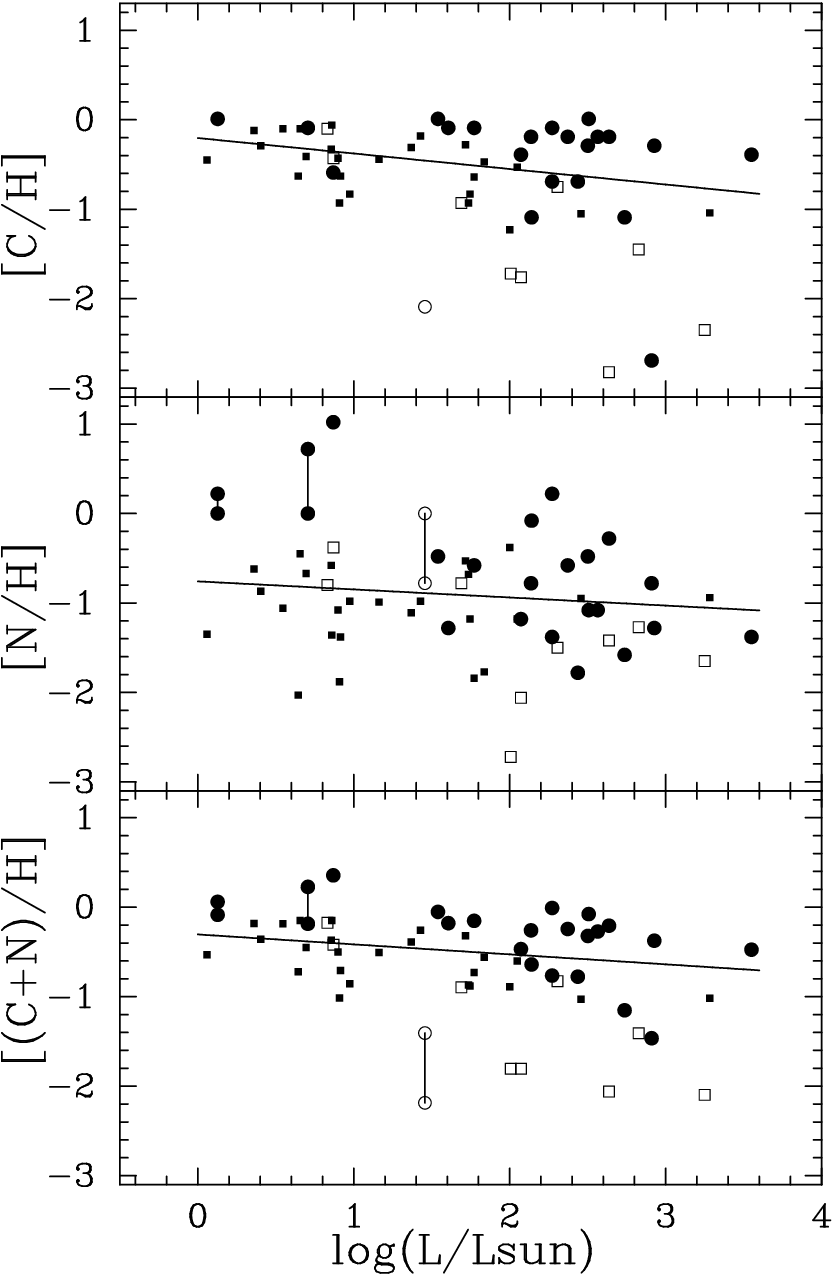} 
\caption[]{[C/H] (top), [N/H] (middle) and [(C+N)/H] (bottom) as a
function of luminosity for Ba-enhanced CEMP stars (filled symbols) and
Ba-normal CEMP stars (open symbols). The solid lines indicate the
least square fit to the data of Ba-enhanced CEMP stars. In the fitting
to the [N/H] and [(C+N)/H] data, the stars for which only upper limits
of N abundances are determined are ignored. Note that,
on average, there exists an alteration of the surface abundances of C
and C+N by about 0.6~dex, as one progresses from relatively unevolved
to evolved stars at higher luminosity.}
\label{fig:chcnhl}
\end{figure} 

\begin{figure} 
\includegraphics[width=8.5cm]{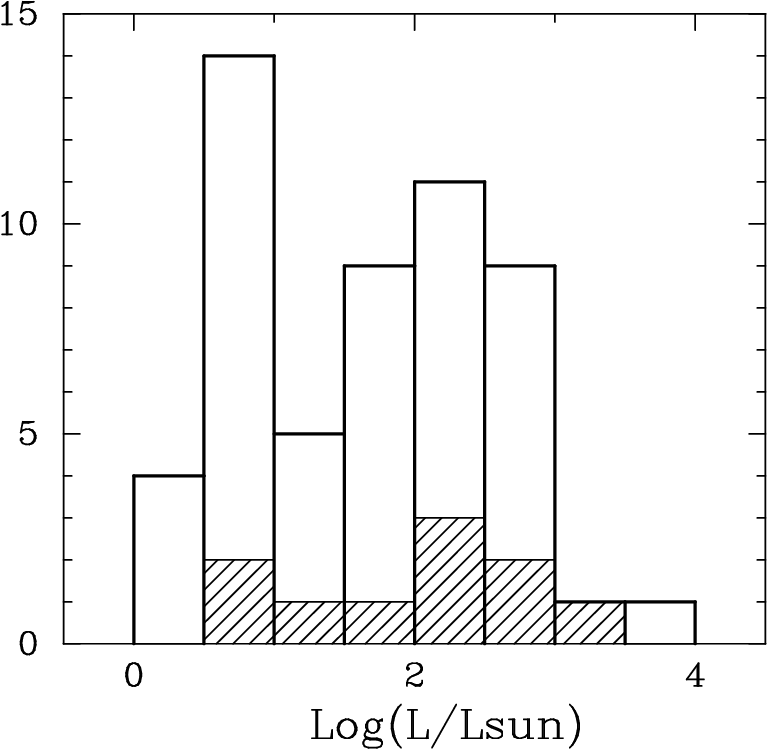} 
\caption[]{The same as Figure~\ref{fig:histfe}, but for the luminosity
  distribution for the CEMP stars in our expanded sample.  The
  distributions of luminosity appear similar for the Ba-enhanced and
  Ba-normal stars (see text).}\label{fig:histl}
\end{figure} 

\begin{figure} 
\includegraphics[width=8.5cm]{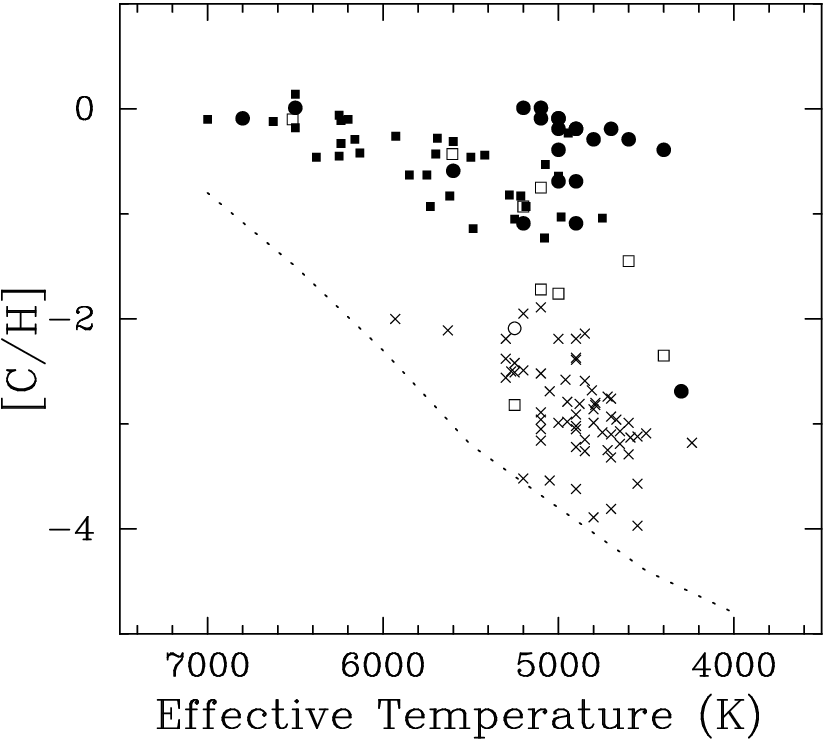} 
\caption[]{The [C/H] ratio as a function of effective temperature for
stars in our expanded sample. The circles and squares indicate CEMP
stars identified by the present and previous work, respectively. The
filled symbols are Ba-enhanced CEMP stars, while the open symbols are
Ba-normal CEMP stars. The cross indicates non carbon-enhanced stars.
The dotted line indicates the value of [C/H] for which the depth of
the CH G band reaches to about 2~\% below the local continuum, as a
function of temperature, estimated from synthetic spectra.  Stars with
[C/H] below this line would not be detectable in high-resolution
spectra of the quality used for this and most previous studies (see
text).  }\label{fig:chteff}
\end{figure} 

\begin{figure} 
\includegraphics[width=15cm]{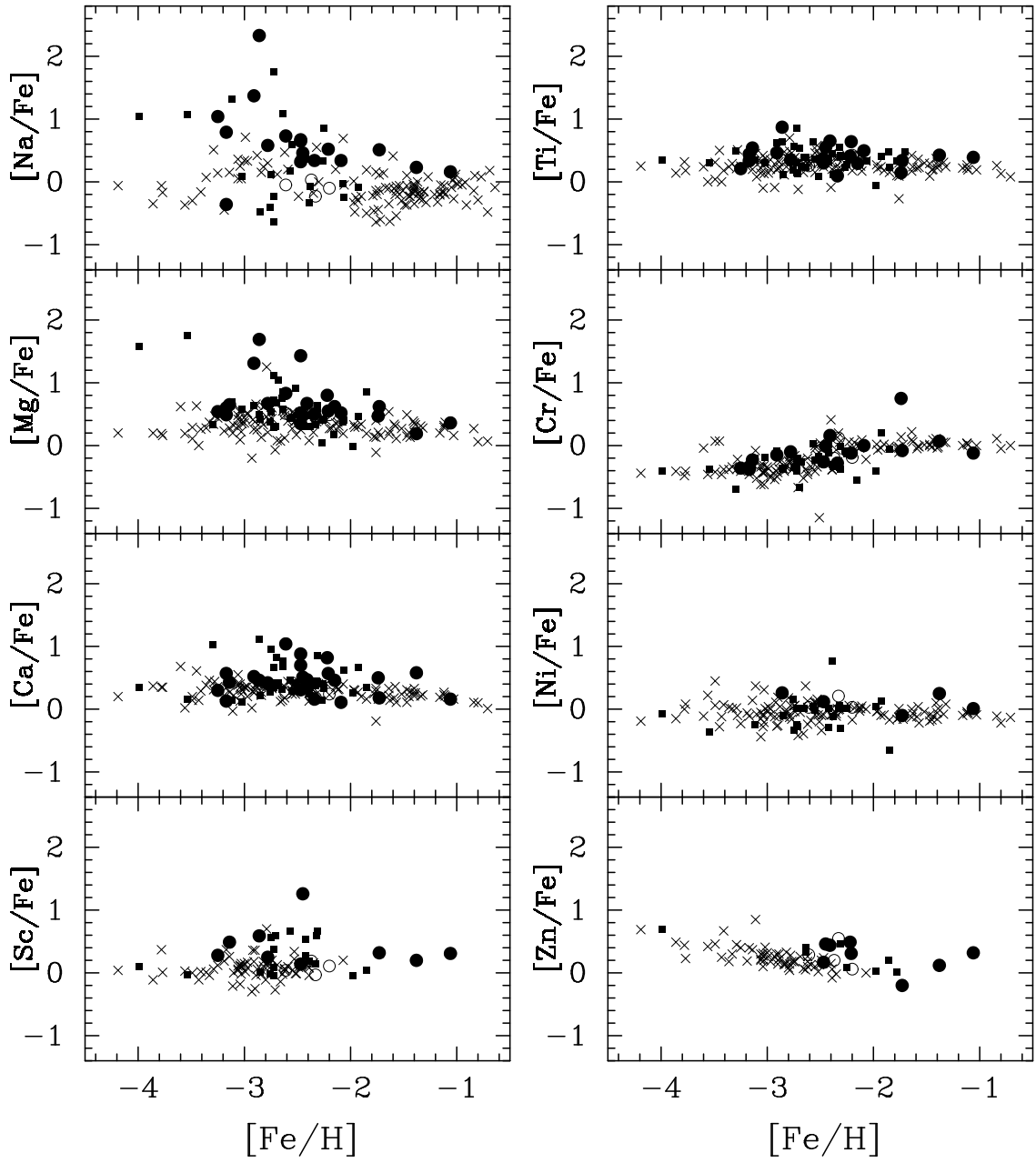} 
\caption[]{Abundance ratios of eight elements as functions of [Fe/H] for stars
in our expanded sample. The symbols are the same as in Figure~\ref{fig:bafe}.}
\label{fig:aball}
\end{figure} 

\begin{figure} 
\includegraphics[width=8.5cm]{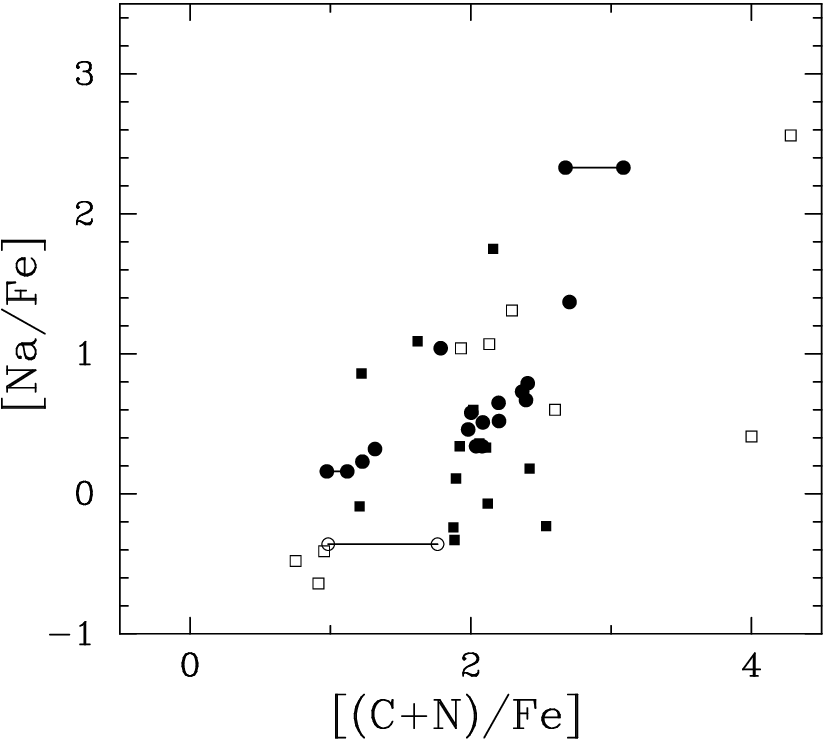} 

\caption[]{[Na/Fe] ratios as a function of [(C+N)/Fe] for stars in our
expanded sample. The filled and open circles indicate Ba-enhanced and
Ba-normal CEMP stars studied by the present work, respectively, while
the filled and open squares exibit Ba-enhanced and Ba-normal CEMP
stars previously studied, respectively.  }

\label{fig:cnna}
\end{figure} 

\end{document}